\documentclass[pra,letterpaper,aps,10pt,superscriptaddress,twocolumn,floatfix,showpacs]{revtex4-1}
\usepackage{graphicx}
\usepackage{amsmath}
\usepackage{amsfonts}
\usepackage{amssymb}
\usepackage{epsfig}
\usepackage{epstopdf}
\DeclareGraphicsExtensions{.pdf,.eps,.png,.jpg,.mps}
\usepackage[pdftex]{color}
\usepackage{amsmath,graphicx,amssymb,braket,xcolor,subfigure,upgreek}
\usepackage[colorlinks, linkcolor=blue, citecolor=blue, urlcolor=blue, breaklinks=true]{hyperref}
\usepackage{microtype}
\usepackage{bbm}
\usepackage{color}

\newcommand{\fref}[1]{Fig.~\ref{#1}}
\newcommand{\fsref}[1]{Figs.~\ref{#1}}
\newcommand{\ts}[1]{_\text{#1}}

\newcommand{\A}{\hat{A}}
\newcommand{\B}{\hat{B}}
\newcommand{\hatS}{\hat{S}}

\begin{document}

\title{Enhanced collective Purcell effect of coupled quantum emitter systems}
\author{D.~Plankensteiner}
\affiliation{Institut f\"ur Theoretische Physik, Universit\"at Innsbruck, Technikerstr. 21a, A-6020 Innsbruck, Austria}
\author{C.~Sommer}
\affiliation{Max Planck Institute for the Science of Light, Staudtstra{\ss}e 2,
D-91058 Erlangen, Germany}
\author{M.~Reitz}
\affiliation{Max Planck Institute for the Science of Light, Staudtstra{\ss}e 2,
D-91058 Erlangen, Germany}
\author{H.~Ritsch}
\affiliation{Institut f\"ur Theoretische Physik, Universit\"at Innsbruck, Technikerstr. 21a, A-6020 Innsbruck, Austria}
\author{C.~Genes}
\affiliation{Max Planck Institute for the Science of Light, Staudtstra{\ss}e 2,
D-91058 Erlangen, Germany}
\date{\today}

\begin{abstract}
Cavity-embedded quantum emitters show strong modifications of free space radiation properties such as an enhanced decay known as the Purcell effect. The central parameter is the cooperativity $C$, the ratio of the square of the coherent cavity coupling strength over the product of cavity and emitter decay rates. For a single emitter, $C$ is independent of the transition dipole moment and dictated by geometric cavity properties such as finesse and mode waist. In a recent work [Phys. Rev. Lett. 119, 093601 (2017)] we have shown that collective excitations in ensembles of dipole-dipole coupled quantum emitters show a disentanglement between the coherent coupling to the cavity mode and spontaneous free space decay. This leads to a strong enhancement of the cavity cooperativity around certain collective subradiant antiresonances. Here, we present a quantum Langevin equations approach aimed at providing results beyond the classical coupled dipoles model. We show that the subradiantly enhanced cooperativity imprints its effects onto the cavity output field quantum correlations while also strongly increasing the cavity-emitter system's collective Kerr nonlinear effect.
\end{abstract}


\maketitle

\section{Introduction}
The decay rate of a quantum emitter placed in an optical resonator can be strongly modified from its bare free space value. The effect stems from the cavity-induced modification of the optical density of states around the emitter's transition frequency. This was predicted by Purcell in $1946$~\cite{purcell1946spontaneous} and measured in various systems employing Fabry-Perot optical cavities, plasmonic modes, microwave cavities, etc.\  \cite{demartini1987anomalous,kreuter2004spontaneous,haase2006detecting}. This indicates the possibility of modifying other properties of materials by dressing them with strongly confined resonant optical fields. For example, at the level of single molecules, the Purcell effect has been employed to controllably tailor the ratio of radiative decay rates from excited zero-phonon electronic states to different ground-state vibrational sublevels, thus enhancing the quantum efficiency~\cite{wang2019turning}. Experimental and theoretical efforts on the collective strong coupling with organic molecules have shown strong modifications of energy and charge transport~\cite{orgiu2015conductivity,schachenmayer2015cavity,feist2015extraordinary,hagenmuller2017cavity,hagenmuller2018cavity}, F\"{o}rster resonance energy transfer~\cite{zhong2016non,zhong2017energy}, chemical reaction rates~\cite{hutchinson2012modifying, galego2016suppressing}, etc.

It has been recently predicted~\cite{plankensteiner2017cavity} that the collective dynamics of $N$ interacting quantum emitters in the bad cavity regime exhibits a scaling of the cooperativity with the emitter number $N$ beyond the expected linear one. Such a behavior can be tested by scanning a probe laser around the common resonance of the cavity mode and a single collective state of the coupled emitters. A "hole-burning" effect occurs around the common resonance with a frequency window characterized by the collective Purcell-modified emitter decay rate, i.e. the emitters shut off transmission around this frequency. At the single particle level such an antiresonance behavior has been experimentally and theoretically discussed~\cite{rice1996cavity,sames2014antiresonance,wang2017cavity}. At the many particle level, the key point is that closely spaced quantum emitters are subject to intense dipole-dipole interactions leading to collective scattering, as observed experimentally~\cite{pellegrino2014observation} and theoretically discussed mostly in one- and two-dimensional geometries~\cite{pellegrino2014observation,bettles2015cooperative,bettles2016cooperative,bettles2016enhanced,shahmoon2017cooperative,perczel2017topological}. Assuming uniform illumination of the dense ensemble (characterized by particle separations smaller than the wavelength of incoming light), carefully chosen lattice constants can ensure that collective subradiant states are addressed that can efficiently reflect light. Theoretical proposals have been directed towards engineering metamaterials with controlled transparency~\cite{jenkins2013metamaterial}, the study of collective motion of atomically thin metamaterials and their interactions with light (opto-nanomechanics)~\cite{botter2013collectivemotion,shahmoon2018collective, shahmoon2018optomechanics} or the enhancement of nonlinear effects~\cite{habibian2011quantum,wild2018quantum}. Engineered interactions via common coupling of emitters to guided modes of a two-dimensional photonic crystal allowed for the theoretical study of topological quantum optics~\cite{perczel2017topological}. In one and two dimensions, collective subradiant states have also been studied for the possibility of robust light-storage devices~\cite{plankensteiner2015selective,facchinetti2016storing}.

One of the widely used theoretical approaches (including in Ref.~\cite{plankensteiner2017cavity}) to describe the response of the quantum emitter ensemble to a driving light field is based on a mapping to a classical problem of coupled dipoles. The simplifying assumption is that in the weak excitation regime the emitters behave as classical oscillators. Collective effects such as superradiance and subradiance are indeed recovered in such an approach. For the treatment in Ref.~\cite{plankensteiner2017cavity}, this approach sufficed to give rise to a semianalytical expression of the transmission of light through a cavity containing a collection of interacting emitters; the results indicated a strong modification of the cavity cooperativity around collective antiresonances associated with collective subradiant states. However, questions regarding the quantum effect of subradiance imprinted on the cavity outgoing light were left open. This paper provides an extension to the quantum problem: We focus here on describing the quantum properties of the output cavity fields (transmitted and reflected) as well as of the detected signal. Linearizing the quantum fluctuations around the classical problem allows us to identify regimes of cooperative enhancement of quadrature squeezing and strongly modified signatures in the second-order correlation functions. The treatment of the classical problem beyond the weak excitation regime also allows an analysis of the collective Kerr effect: Around subradiant antiresonances, the third-order nonlinear response of the system is greatly enhanced even for limited numbers of emitters.

The paper is organized as follows: In Sec.~\ref{cavitydynamics}, we introduce the full model of $N$ quantum emitters interacting with a cavity field both within the master equation formalism as well as quantum Langevin equations (QLEs). We proceed by justifying the linear approximation and deriving coupled equations of motion for classical averages as well as fluctuation operators. In Sec.~\ref{singleemitter}, we quickly review fundamentals of cavity quantum electrodynamics (cavity QED) with a single emitter such as occurrence of strong coupling, the Purcell effect, and antiresonances. We then derive the classical response in reflection and transmission for two-sided cavities. We introduce operators for the detected signal and provide a formalism for computing variances and correlations for intracavity, outgoing, and detected fields. In Sec.~\ref{freespace}, we describe some fundamental aspects of vacuum-coupled quantum emitter ensembles exhibiting subradiance and superradiance and investigate some of their entanglement properties. Finally, in Sec.~\ref{collective}, we present the dynamics of coupled emitter ensembles inside a common cavity mode, exhibiting a modified collective Purcell effect, and analytically derive cavity transmission properties, equations of motion for the fluctuation operators, and the modification of the collective third-order nonlinearity.

\begin{figure}[t]
\includegraphics[width=0.8\columnwidth]{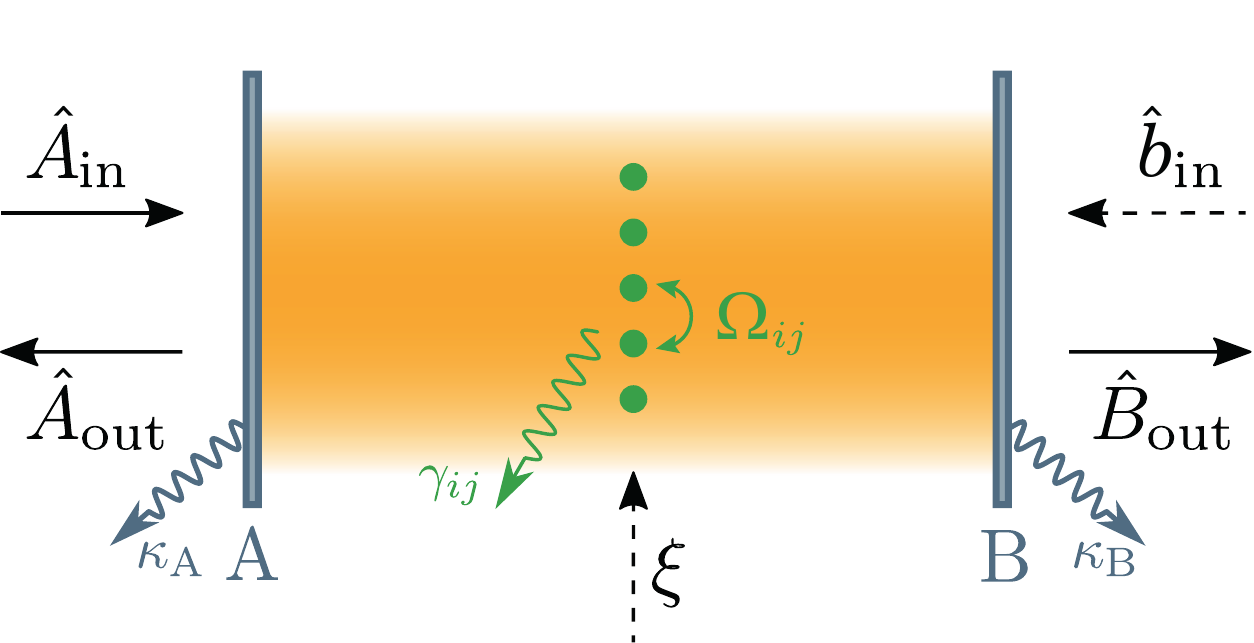}
\caption{\emph{Model schematic}. Optical cavity containing $N$ coupled, closely spaced quantum emitters. The vacuum modes (not supported by the cavity) mediate dipole-dipole interactions with strength $\Omega_{ij}$ and induce collective decay with $\gamma_{ij}$. Losses through the mirrors occur at rates $\kappa\ts{A}$ (left mirror) and $\kappa\ts{B}$ (right mirror). The cavity is pumped through the left mirror with nonzero amplitude $\A \ts{in}$ while zero-average noise is entering via the right mirror as $\hat{b} \ts{in}$. Transmission and reflection are measured by detecting outgoing nonzero-average operators $\A\ts{out}$ and $\B\ts{out}$.}
\label{fig1}
\end{figure}

\section{Cavity dynamics of coupled quantum emitters}
\label{cavitydynamics}
We consider an ensemble of $N$ quantum emitters each with a ground state $\ket{g}_j$ and an excited state $\ket{e}_j$ (resonance frequency $\omega\ts{e}$) located at $\textbf{r}_j$. The corresponding raising and lowering operators are denoted by $\hat S_j^\dagger$ and $\hat S_j$, respectively. The emitters are placed within a plane orthogonal to the cavity axis and inside the waist of a cavity mode at frequency $\omega_\text{c}$  (see~\fref{fig1}). The cavity is laser driven at frequency $\omega_{\ell}$ with power $\cal P$ through the left mirror. The coherent cavity mode dynamics are described by the Hamiltonian (in a frame rotating at $\omega_\ell$),
\begin{align}
H \ts{cav} = -\hbar\Delta\ts{c}\A^\dag \hat{A} + i\hbar\eta\left(\A^\dag - \A\right),
\end{align}
where $\Delta\ts{c}=\omega_{\ell}-\omega\ts{c}$ and $\eta=\sqrt{2\cal {P} \kappa\ts{A}/(\hbar\omega_{\ell})}$. The cavity damping rate is $\kappa=(\kappa\ts{B}+\kappa\ts{A})/2$ (encompassing losses via both left and right mirrors) and occurs via the collapse operator $\A$ contained in the Lindblad term, $\mathcal{L}\ts{c}[\rho] = \kappa\left(2\A\rho \A^\dag - \A^\dag \A \rho - \rho \A^\dag \A\right)$.

In the single-mode limit, the emitter-cavity interaction is described by the Tavis-Cummings Hamiltonian
\begin{align}
H\ts{int} = \hbar\sum_j g_j \left(\A^\dag \hatS_j +  \A \hatS_j^\dag \right),
\end{align}
where each of the emitters couples to the cavity mode with a distinct rate $g_j$ which depends on the emitter positions as well as the cavity mode profile.
At dense spacing ($d:=|\textbf{r}_j - \textbf{r}_{j+1}| < \lambda\ts{e}$), one has to account for the vacuum-mediated emitter-emitter interactions via the fields they emit due to their transition dipole moments $\boldsymbol{\mu}$ (assuming all dipole moments to be identical). The total emitter Hamiltonian includes a free part plus the collective coherent dipole-dipole interactions
\begin{align}
H\ts{e} = -\hbar\sum_{j}\Delta\ts{e} \hatS_j^\dag \hatS_j + \hbar\sum_{j\neq k}\Omega_{jk} \hatS_j \hatS_k^\dag ,
\end{align}
where $\Delta\ts{e} = \omega_{\ell}-\omega\ts{e}$. The dipole-dipole interactions governed by the frequencies $\Omega_{ij}$ are derived by eliminating the degrees of freedom for the surrounding vacuum modes excluding the single mode supported by the cavity (see Appendix~\ref{E-field-derivation}). Doing so additionally leads to dissipation of the emitters in the form of collective decay with rates $\gamma_{ij}=\gamma h_{ij}$ (where $h_{ij}$ is maximally unity for $i=j$) that are described by the Lindblad term~\cite{lehmberg1970radiation}
\begin{align}
\mathcal{L}\ts{e} [\rho] = \sum_{j,k} \gamma h_{jk}\left( 2 \hatS_j \rho \hatS_k^\dag - \hatS_j^\dag \hatS_k \rho - \rho \hatS_j^\dag \hatS_k\right).
\end{align}
While full numerical simulations for moderate numbers of quantum emitters can be carried out based on the master equation description, $\partial_t \rho(t)=i [\rho(t),H]/\hbar+\mathcal{L}[\rho(t)]$, we move to an equivalent quantum Langevin approach (see Appendix~\ref{sec:App.QLEs}), which allows for the derivation of analytical results,
\begin{widetext}
\begin{subequations}
\label{N_QLE_nl}
\begin{align}
\dot \A &= -(\kappa -i\Delta\ts{c}) \A -i\sum_{j} g_j \hatS_j + \eta +  \sqrt{\kappa\ts{A}}\hat{a}\ts{in}+\sqrt{\kappa\ts{B}}\hat{b}\ts{in},\\
\dot \hatS_j  &=-(\gamma -i\Delta\ts{e})\hatS_j + i g_j \A \hatS_{j}^z + \sum_{k\neq j} \left(i\Omega_{jk} + \gamma_{jk}\right)\hatS_j^z \hatS_k -\sqrt{2\gamma}\bar{\xi}_j(t),
\\
\dot \hatS_j^z &= -2\gamma (\hatS_j^z +1 ) + 2ig_j\left(\A^\dag \hatS_j - \hatS_j^\dag \A\right) - \sum_{k\neq j}2\gamma_{jk}\left(\hatS_j^\dag \hatS_k + \hatS_k^\dag \hatS_j\right) + \sqrt{2\gamma}\bar{\xi}_j^z(t).\end{align}
\end{subequations}
\end{widetext}

The convention in this paper is that nonzero average operators are denoted by capital letters while lowercase letters denote fluctuation operators. The left mirror allows for a nonzero average input $\A\ts{in}=\eta/\sqrt{\kappa\ts{A}}+\hat{a}\ts{in}$ with the zero-average white-noise term fulfilling $\langle \hat{a}\ts{in}(t) \hat{a}\ts{in}^\dag (t')\rangle=\delta(t-t')$ (while all other correlations vanish). The right mirror allows for white noise only with all correlations vanishing except for $\langle \hat{b}\ts{in}(t) \hat{b}\ts{in}^\dag (t')\rangle=\delta(t-t')$. One can also define an effective input operator,
\begin{align}
\hat{C}\ts{in}=\sqrt{\frac{\kappa\ts{A}}{\kappa\ts{A}+\kappa\ts{B}}}\A\ts{in}+\sqrt{\frac{\kappa\ts{B}}{\kappa\ts{A}+\kappa\ts{B}}}\hat{b}\ts{in},
\label{N_QLE_nl_avg}
\end{align}
in terms of which the QLE for the cavity field shows a single compound input noise added as $\hat{c}\ts{in}$ with $\langle \hat{c}\ts{in}(t) \hat{c}\ts{in}^\dag (t')\rangle=\delta(t-t')$ and $\hat{c}\ts{in}=\hat{C}\ts{in}-\eta/\sqrt{2\kappa}$. On the quantum emitter side we have defined effective noise operators affecting the emitters (see Appendix~\ref{sec:App.QLEs} for more details) $\bar{\xi}_j(t) = \hatS_j^z \xi_j(t)$ and $\bar{\xi}_j^z(t) = 2\left(\hatS_j^\dag\xi_j(t) + \xi_j^\dag(t)\hatS_j\right)$.  In the absence of classical drive terms for the quantum emitters, the noises are zero-average and $\delta$ correlated in time; however, as the emitters are placed in the near field of their neighbors, spatial correlations are included in the pairwise decay terms, i.e., $\braket{\xi_i(t)\xi_j^\dag(t')}=h_{ij}\delta(t-t')$. A linearization procedure can be applied around the average values ($\alpha=\langle \A \rangle$, $\beta_j=\langle \hatS_j \rangle$ and $z_j=\braket{\hatS_j^z}$), introducing zero-average fluctuation operators $\hat{a}=\A-\alpha$, $\hat{\sigma}_j=\hat{S}_j-\beta_j$, and $\hat{\sigma}_j^z = \hat{S}_j^z - z_j$, respectively. We then proceed by neglecting products of fluctuation operators. This allows us to derive two distinct sets of equations, one for the classical averages (which still includes non-linear behavior as long as we keep the equation for the population inversion) and one set for the fluctuation operators (linearized). For the classical averages, we find
\begin{subequations}
\begin{align}
\dot \alpha &= -(\kappa -i\Delta\ts{c}) \alpha -i\sum_{j} g_j \beta_j + \eta,\\
\dot \beta_j  &=-(\gamma -i\Delta\ts{e})\beta_j + i g_j \alpha z_j + \sum_{k\neq j} \left(i\Omega_{jk} + \gamma_{jk}\right) z_j \beta_k,\\
\dot z_j &= -2\gamma \left(z_j + 1\right) + 2ig_j\left(\alpha^*\beta_j - \beta_j^*\alpha\right) +
\notag \\
&- 4\sum_{k\neq j}\gamma_{jk}\Re\left\{\beta_j^*\beta_k\right\}.
\label{N_QLE_lin_avg}
\end{align}
\end{subequations}
Note that, in this limit, we can express the inversion average as $z_j = \braket{2\hat{S}_j^\dag \hat{S}_j - 1} \approx 2|\beta_j|^2 - 1$ as a second-order perturbation in $\eta$. Next-order terms, stemming from two fluctuation operators averages such as $\langle \hat{\sigma}_j^{\dagger}\hat{\sigma}_j\rangle$, are already fourth order corrections in $\eta$.
We can then write QLEs for the quantum fluctuations of all operators,
\begin{widetext}
\begin{subequations}
\label{N_QLE_lin}
\begin{align}
\dot{\hat{a}} &= -\left(\kappa - i\Delta\ts{c}\right) \hat{a} - i\sum_jg_j\hat{\sigma}_j +\sqrt{\kappa\ts{A}}\hat{a}\ts{in}+\sqrt{\kappa\ts{B}}\hat{b}\ts{in},
\\
\dot{\hat{\sigma}}_j &= -\left(\gamma - i\Delta\ts{e}\right)\hat{\sigma}_j + ig_j\left(z_j \hat{a} + \alpha\hat{\sigma}_j^z\right) +
\sum_{k\neq j}\left(i\Omega_{jk} + \gamma_{jk}\right)\left(z_j\hat{\sigma}_k + \beta_k\hat{\sigma}_j^z\right) - \sqrt{2\gamma}\bar{\xi}_j(t),
\\
\dot {\hat{\sigma}}_j^z &= -2\gamma\hat{\sigma}_j^z + 2ig_j\left(\alpha^*\hat{\sigma}_j + \beta_j \hat{a}^\dag - \alpha\hat{\sigma}_j^\dag - \beta_j^* \hat{a}\right) - 2\sum_{k\neq j}\gamma_{jk}\left(\beta_j^*\hat{\sigma}_k + \beta_k\hat{\sigma}_j^\dag + \text{H.c.}\right) + \sqrt{2\gamma}\bar{\xi}_j^z(t).
\label{N_QLE_lin_end}
\end{align}
\end{subequations}
\end{widetext}
Let us now discuss the correlations of the emitter noise terms. Assuming the environment for the emitter input noise to be in a vacuum state, the effective noise terms are also of zero average. However, they have the following non-vanishing correlations,
\begin{subequations}
\label{noise-corr}
\begin{align}
\label{noise-corr-1}
\braket{\bar{\xi}_j(t)\bar{\xi}_{k}^\dag(t')} &=
  \begin{cases}
    \delta(t-t'),       & \quad \text{if } j=k\\
    h_{jk}z_jz_k\delta(t-t'),  & \quad \text{if } j\neq k
  \end{cases}
\end{align}
\begin{align}\label{noise-corr-2}
 \braket{\bar{\xi}_j^z(t)\bar{\xi}_{k}^z(t')} &=
  \begin{cases}
    2\left(z_j + 1\right)\delta(t-t'),       & \quad \text{if } j=k\\
   4 h_{jk}\beta_j^*\beta_k\delta(t-t'),  & \quad \text{if } j\neq k
  \end{cases}
\end{align}
\begin{align}
\braket{\bar{\xi}_j^z(t)\bar{\xi}_k^\dag(t')}  &=
  \begin{cases}
    -2\beta_j^*\delta(t-t'),       & \quad \text{if } j=k\\
     2h_{jk}z_k\beta_j^*\delta(t-t'),  & \quad \text{if } j\neq k
  \end{cases}
\label{noise-corr-end}
\end{align}
\end{subequations}
and $\braket{\bar{\xi}_j(t)\bar{\xi}_k^z(t')} = \braket{\bar{\xi}_k^z(t')\bar{\xi}_j^\dag(t)}^*$.

\begin{figure*}[t]
\includegraphics[width=1.95\columnwidth]{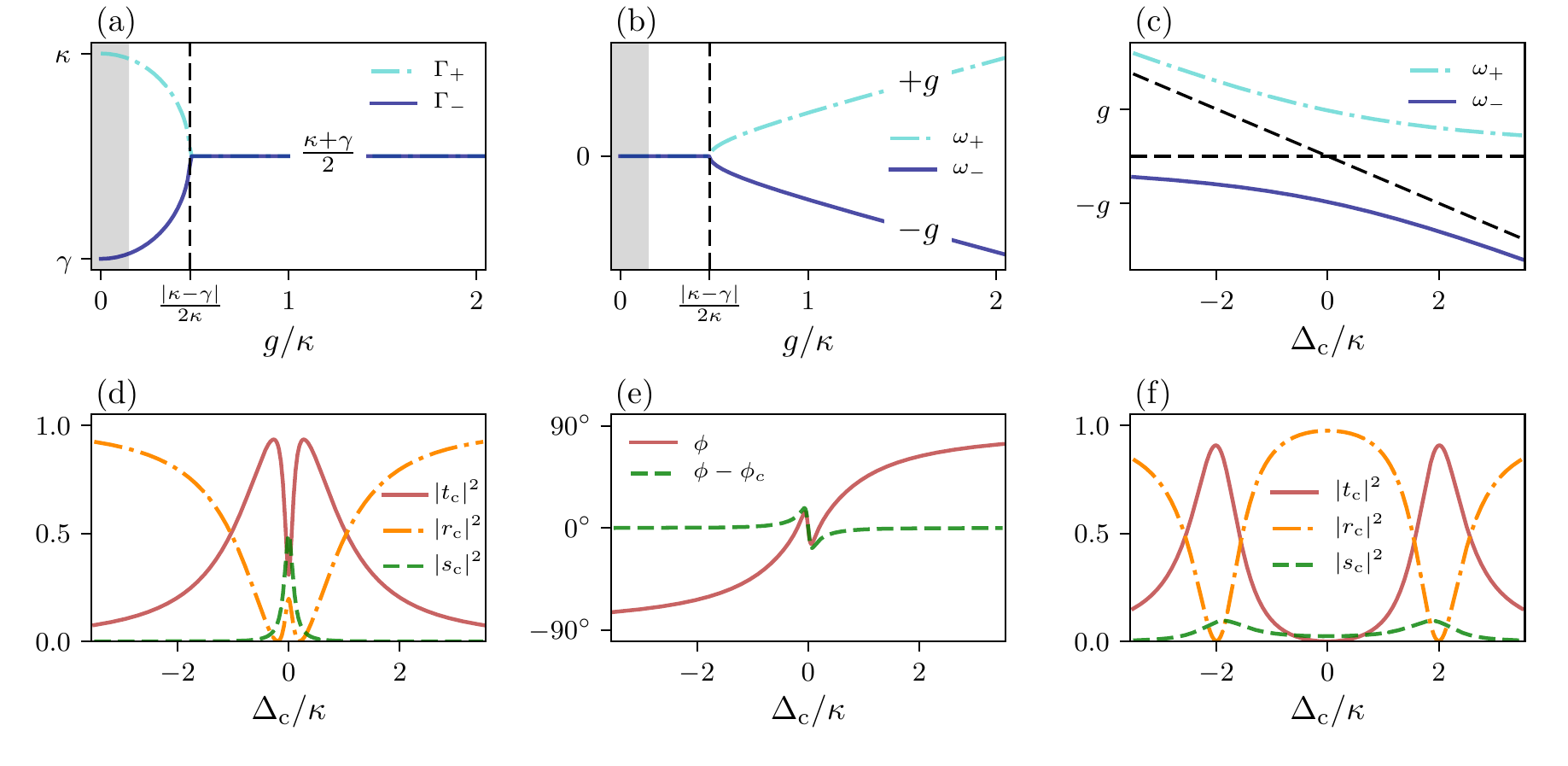}
\caption{\emph{Single emitter cavity mode hybridization}. Plot of hybrid cavity-emitter decay rates (a) and frequencies (b) when sweeping the coupling $g$ past the strong coupling onset point of $g=|(\kappa-\gamma)/2|$ (in the resonant case). The gray region shows the weak coupling but strong cooperativity regime where the Purcell effect shows up as a modification of the emitter's radiative rate. (c) Standard picture of avoided resonances in the strong-coupling regime when the cavity detuning is scanned. (d) Intensity of transmitted, reflected, and scattered fields of an emitter-cavity system in the Purcell regime (antiresonance regime) for a laser scan around the resonance. The parameters are $g=\kappa/5$, $\gamma=\kappa/20$. (e) Cavity phase shift and emitter-only induced phase shift in the same regime. (f) Cavity response in transmission and reflection as well as scattered field showing the signature of polaritons in the strong coupling regime ($g=2\kappa$).}
\label{fig2}
\end{figure*}

\section{Single-emitter antiresonance spectroscopy}
\label{singleemitter}

Let us first fully analyze the emitter-cavity mode hybridization by solving Eqs.~\eqref{N_QLE_nl} for a single emitter. Steady-state solutions for the operator averages already suffice to provide an overview of effects such as cavity strong coupling (occurrence of polaritons), antiresonances, and the Purcell modification of the emitter's decay rate. We then make the connection between the intracavity dynamics and the amplitude and phase transmission/reflection for asymmetric two-sided cavities. In the next step we describe the quantum properties of the field inside the cavity and of the output fields (in transmission/reflection). By assuming a particular detection scheme which allows us to define nondimensional operators for the detected field, we analyze the connection between the detected signal and the continuous output fields. Finally, we compute the next-order correction to the steady-state solution to derive the scaling of the system's Kerr nonlinearity.

\subsection{Regimes of interaction}
The classical equations of motion for the cavity field amplitude and the dipole of the quantum emitter are sufficient to characterize the different regimes of interaction inside the optical cavity,
\begin{subequations}
\label{single_HE}
\begin{align}
\dot{\alpha} &= -\left(\kappa -i\Delta\ts{c}\right)\alpha -ig\beta+ \eta,
\label{single_HE1}
\\
\dot{\beta} &= -(\gamma-i\Delta\ts{e})\beta - i g \alpha.
\label{single_HE2}
\end{align}
\end{subequations}
We denoted by $\kappa$ the effective decay rate via both mirrors $\kappa=(\kappa_A+\kappa_B)/2$. The diagonalization of the above equations (under resonance conditions, i.e. fixed $\Delta\ts{c}=\Delta\ts{e}=0$) leads to the hybridized decay rates and frequencies,
\begin{subequations}
\begin{align}
\Gamma_\pm &= \frac{\kappa + \gamma}{2} \pm \Re\left\{\sqrt{\left(\frac{\kappa - \gamma}{2}\right)^2 - g^2}\right\},
\\
\omega_\pm &= \pm\Im\left\{\sqrt{\left(\frac{\kappa - \gamma}{2}\right)^2 - g^2}\right\}.
\end{align}
\end{subequations}
The threshold $g>|(\kappa-\gamma)/2|$ indicates the onset of the strong coupling regime where the two frequencies combine into distinct polariton branches. Far above this threshold the polaritons are symmetrically displaced by $\pm g$ from the original energies [see \fref{fig2}(b)]. The decay rates show a different behavior as they already hybridize before the onset of strong coupling and ultimately reach the algebraic average $(\kappa+\gamma)/2$. We will be mostly interested in the weak coupling regime, highlighted in gray in \fref{fig2}(a) and (b), where for $\gamma\ll \kappa$, a strong modification of the emitter bare decay rate by a factor $1+C$ occurs [where $C$ is the cooperativity defined as $C=g^2/(\kappa\gamma)$]. This is the Purcell effect and one can cast the Purcell factor~\cite{purcell1946spontaneous} given by $F\ts{p}=6\pi c^3Q/(\omega\ts{e}^3 V)$ in terms of the cooperativity parameter. Using the definition of the quality factor $Q=\omega\ts{c}/\kappa$, the dipole coupling strength $g = \mu\sqrt{\omega\ts{c}/(2\epsilon_0\hbar V)}$ and the free space decay rate $\gamma = \omega\ts{e}^3\mu^2/(3\pi\hbar c^3\epsilon_0)$, we can express the Purcell factor as $F\ts{p} = 4C$.

\subsection{Antiresonance: Transmission, reflection and absorption}

Assuming steady state, we set the derivatives to zero in Eqs.~\eqref{single_HE} and obtain
\begin{subequations}
\begin{align}
\alpha &= \frac{-ig\beta + \eta}{ \kappa-i\Delta\ts{c}},
\\
\beta &= \frac{-ig \alpha}{\gamma-i\Delta\ts{e}}.
\end{align}
\end{subequations}
Under the considered approximations, the dipole responds linearly to the intra-cavity field; the cavity field in turn is the result of interference between the pump signal and the dipole re-radiated amplitude. Solving the above equations we find
\begin{subequations}
\begin{align}
\alpha &= \frac{\eta \left(\gamma-i\Delta\ts{e}\right)}{\left(\kappa-i\Delta\ts{c} \right)\left(\gamma-i\Delta\ts{e}\right) + g^2},
\\
\label{dipolelinear}
\beta &= \frac{-ig\eta}{\left(\kappa-i\Delta\ts{c} \right)\left(\gamma-i\Delta\ts{e}\right) + g^2}.
\end{align}
\end{subequations}
The cavity output signal consists of three parts: the reflected ($r\ts{c}$), transmitted ($t\ts{c}$) and scattered ($s\ts{c}$) field. The latter is the field leaking out of the sides of the cavity due to spontaneous decay of the emitter. In order to investigate these three parts, we make use of the input-output relations written separately at both left and right mirrors,
\begin{subequations}
\label{input-output}
\begin{align}
\A\ts{in} + \A\ts{out} &= \sqrt{\kappa\ts{A}} \A,
\\
\hat{b}\ts{in} + \B\ts{out} &=  \sqrt{\kappa\ts{B}}\A.
\end{align}
\end{subequations}
As specified above, driving is done through the left mirror such that $\langle \A\ts{in}\rangle =\eta/\sqrt{\kappa\ts{A}}$. Averaging of the equations above thus leads to
\begin{subequations}
\begin{align}
\braket{\B\ts{out}} &=  \sqrt{\kappa\ts{B}}\alpha,
\\
\eta/\sqrt{\kappa\ts{A}} + \braket{\A\ts{out}} &=  \sqrt{\kappa\ts{A}}\alpha.
\end{align}
\end{subequations}
The amplitude transmission coefficient $t\ts{c}$ and the reflection coefficient $r\ts{c}$, respectively, are then given by
\begin{subequations}
\begin{align}
t\ts{c} &= \frac{\braket{\B\ts{out}}}{\braket{\A\ts{in}}} =  \frac{\sqrt {\kappa\ts{A}\kappa\ts{B}}}{\eta}\alpha,
\\
r\ts{c} &= \frac{\braket{\A\ts{out}}}{\braket{\A\ts{in}}} = \frac{\kappa\ts{A}}{\eta}\alpha - 1 = t\ts{c}\sqrt{\frac{\kappa\ts{A}}{\kappa\ts{B}}} - 1.
\end{align}
\end{subequations}
While generally the cavity properties strongly depend on the mirror asymmetry, let us focus on a perfectly balanced cavity where $\kappa=\kappa\ts{A}=\kappa\ts{B}$ and express the complex transmission amplitude as
\begin{align}
\label{trans}
t\ts{c} &= \frac{\kappa}{\kappa-i\Delta\ts{c}+g^2/(\gamma-i\Delta\ts{e} )},
\end{align}
while the reflectivity is immediately derived as $r=-1+t$. This expression already contains the phenomenon of emitter antiresonances \cite{plankensteiner2017cavity,alsing1992suppression,sames2014antiresonance}, where the resonantly driven dipole oscillates in a way to counteract the cavity drive leading to a local minimum of transmission~\cite{zippilli2004suppressing}.

The respective intensities are given by the absolute squares of the complex coefficients. We note, that it is possible to write down a similar input-output relation for the scattered field in the linearized regime. However, for more general purposes, one can use the fact that the sum of all the intensities has to be conserved, namely $|r\ts{c}|^2 + |t\ts{c}|^2 + |s\ts{c}|^2 = 1$. This gives the scattered intensity $|s\ts{c}|^2 = 1 - |t\ts{c}-1|^2 - |t\ts{c}|^2 = 2\left(\Re\left\{t\ts{c}\right\} - |t\ts{c}|^2\right)$. At resonance ($\Delta\ts{c}=\Delta\ts{e}=0$), we can express all the intensities in terms of the cooperativity,
\begin{subequations}
\begin{align}
|t\ts{c}|^2 &= \frac{1}{\left(1+C\right)^2},
\\
|r\ts{c}|^2 &= \frac{C^2}{\left(1+C\right)^2},
\\
|s\ts{c}|^2 &= \frac{2C}{\left(1+C\right)^2}.
\end{align}
\end{subequations}
An interesting point here is the scaling of the scattered field with the cooperativity. Namely, not only does it vanish for small cooperativity (where the emitter is simply never excited and thus cannot scatter), but also for $C\gg 1$ the radiation to the side is suppressed. Since the transmission vanishes as well in this regime, the entire input field is reflected [see \fref{fig2}(f)].

The phase shift of the field that passes through the cavity is given by the transmission coefficient as
\begin{align}
\phi &= \text{Arg}(t\ts{c}) = \arctan\left(\frac{\Im\{t\ts{c}\}}{\Re\{t\ts{c}\}}\right).
\end{align}
While this corresponds to the phase shift caused by the hybrid system, in the resonant case one can approximate the phase shift caused only by the emitter by subtracting the empty-cavity response $\phi\ts{c} = \arctan\left(\Delta\ts{c}/\kappa\right)$ [see \fref{fig2}(e)].

\subsection{Intracavity steady state}

In a first step we will write the QLEs in the single emitter case and solve for the intra-cavity fluctuation operators in steady state. For $N=1$, Eqs.~\eqref{N_QLE_lin} reduce to
\begin{subequations}
\begin{align}
\dot{\hat{a}} &= -\left(\kappa - i\Delta\ts{c}\right)\hat{a} - ig\hat{\sigma} + \sqrt{\kappa} \hat{a}\ts{in}(t) + \sqrt{\kappa} \hat{b}\ts{in}(t),
\\
\dot{\hat{\sigma}} &= -\left(\gamma - i\Delta\ts{e}\right)\hat{\sigma} + ig\left(\alpha\hat{\sigma}^z + z\hat{a}\right) - \sqrt{2\gamma} \bar{\xi}(t),
\\
\dot {\hat{\sigma}}^z &= -2\gamma\hat{\sigma}^z + 2ig\left(\alpha^*\hat{\sigma} + \beta \hat{a}^\dag - \alpha \hat{\sigma}^\dag - \beta^*\hat{a}\right)+
\\
&+ \sqrt{2\gamma}\bar{\xi}^z(t). \notag
\end{align}
\end{subequations}
The correlation functions for the single-emitter input noise are derived from Eqs.~\eqref{noise-corr-1}--\eqref{noise-corr-end} for $N=1$. We proceed by casting the above set of equations in a more convenient matrix form with the following definitions:
\begin{align}
\textbf{v} := \begin{pmatrix}
\hat{a} \\
\hat{a}^\dag \\
\hat{\sigma} \\
\hat{\sigma}^\dag\\
\hat{\sigma}^z
\end{pmatrix},
\quad
\textbf{v}\ts{in} := \begin{pmatrix}
\hat{a}\ts{in} \\
\hat{a}\ts{in}^\dag \\
\hat{b}\ts{in} \\
\hat{b}\ts{in}^\dag \\
\bar{\xi} \\
\bar{\xi}^\dag\\
\bar{\xi}^z
\end{pmatrix}
\end{align}
The system dynamics can then be written as a matrix-vector differential equation,
\begin{align}
\label{QLE_single_vec}
\dot{\textbf{v}} &= \textbf{M}\textbf{v} + \textbf{N}\textbf{v}\ts{in}(t).
\end{align}
The matrix \textbf{M} is a drift matrix that is completely determined by steady-state expectation values,
\begin{widetext}
\begin{align}
\textbf{M} = \begin{pmatrix}
-\left(\kappa - i\Delta\ts{c}\right) & 0 & -ig & 0 & 0 \\
0 & -\left(\kappa + i\Delta\ts{c}\right) & 0 & ig & 0\\
igz & 0 & -\left(\gamma - i\Delta\ts{e}\right) & 0 & ig\alpha \\
0 & -igz & 0 & -\left(\gamma + i\Delta\ts{e}\right) & -ig\alpha^* \\
-2ig\beta^* & 2ig\beta & 2ig\alpha^* & -2ig\alpha & -2\gamma
\end{pmatrix}.
\end{align}
\end{widetext}
The input noise terms are multiplied with the matrix \textbf{N}, which is given by the decay rates for each dissipation channel,
\begin{align}
\textbf{N} = \begin{pmatrix}
\sqrt{\kappa} & 0 &\sqrt{\kappa} & 0 & 0 & 0 & 0\\
0 & \sqrt{\kappa} & 0 & \sqrt{\kappa} & 0 & 0 & 0\\
0 & 0 & 0 & 0 & -\sqrt{2\gamma} & 0 & 0\\
0 & 0 & 0 & 0 & 0 & -\sqrt{2\gamma} & 0\\
0 & 0 & 0 & 0 & 0 & 0 & \sqrt{2\gamma}
\end{pmatrix}.
\end{align}
Formal integration of the fluctuation operators' QLEs, Eq.~\eqref{QLE_single_vec}, leads to
\begin{align}
\textbf{v}(t) &= e^{\textbf{M}t}\textbf{v}(0) + \int_{0}^t dt' e^{\textbf{M}(t-t')}\textbf{N}\textbf{v}\ts{in}(t'),
\end{align}
where the first term is the transient solution. Assuming that the system is stable, i.e. all the eigenvalues of the drift matrix have negative real parts, this solution vanishes at large times and the system reaches a unique steady state independent of the initial conditions. One can then fully analyze the properties of the system in steady state by looking at the fluctuation correlation matrix $\textbf{V} = \braket{\textbf{v}(t)\textbf{v}^\top(t)}$. The correlations of all input noises can be jointly written as $\braket{\textbf{v}\ts{in}(t)\textbf{v}^\top\ts{in}(t')}=\delta(t-t')\textbf{C}$, where the noise correlation matrix is
\begin{align}
\label{defC}
\textbf{C} =
\begin{pmatrix}
0 & 1 & 0 & 0 & 0 & 0 & 0\\
0 & 0 & 0 & 0 & 0 & 0 & 0\\
0 & 0 & 0 & 1 & 0 & 0 & 0\\
0 & 0 & 0 & 0 & 0 & 0 & 0\\
0 & 0 & 0 & 0 & 0 & 1 & -2\beta\\
0 & 0 & 0 & 0 & 0 & 0 & 0\\
0 & 0 & 0 & 0 & 0 & -2\beta^* & 2\left(1+z\right)
\end{pmatrix}.
\end{align}
A diffusion matrix can afterwards be constructed $\textbf{D}=\textbf{N}\textbf{C}\textbf{N}^\top$ and a simplified equation involving only the correlation and the diffusion matrix can be obtained (see details in Appendix \ref{sec:App.Lyapunov}),
\begin{align}
\textbf{M}\textbf{V} &+ \textbf{V}\textbf{M}^\top =  -\textbf{D}.
\end{align}
This is known as the Lyapunov equation and it allows one to derive all two-point operator correlations for the system in the long-time limit (steady-state condition).

\subsection{Output fields}

To derive quantum properties of the field exiting the cavity we use the input-output relations such as the ones in Eq. \eqref{input-output}. All input-output relations combined into a convenient vector form read
\begin{align}
\textbf{v}\ts{out}(t) &= \textbf{N}^\top\textbf{v}(t) - \textbf{v}\ts{in}(t).
\end{align}
In the time domain the output field is not $\delta$ correlated, which makes calculations more cumbersome. However, a transformation to the Fourier domain provides an immediate simplification as the output is obtained as a matrix multiplication of the input, ensuring that the $\delta$ correlations in the frequency domain are still valid. First, we express the output Fourier components in terms of input noise,
\begin{align}
\textbf{v}\ts{out}(\omega) &= \textbf{F}(\omega)\textbf{v}\ts{in}(\omega),
\end{align}
where $\textbf{F}(\omega) := [\textbf{N}^{\top}\left(i\omega \mathbbm{1} - \textbf{M}\right)^{-1}\textbf{N} - \mathbbm{1}]$. This allows one to compute any correlations
\begin{align}
\braket{\textbf{v}\ts{out}(\omega)\textbf{v}^\top \ts{out}(\omega')}=\pmb{\cal{S}}\ts{out}(\omega)\delta(\omega+\omega'),
\end{align}
contained in the frequency spectrum matrix $\pmb{\cal{S}}\ts{out}$ compactly expressed as
\begin{align}
\label{S_out}
\pmb{\cal{S}}\ts{out}(\omega) = \textbf{F}(\omega)\textbf{C} \textbf{F}^{\top}(-\omega).
\end{align}
As we will see in the following, the output spectrum matrix contains all the information required to compute quantum properties such as squeezing of the output field or of the detected field or the variance of the detected photon number.

\begin{figure*}[t]
\includegraphics[width=1.95\columnwidth]{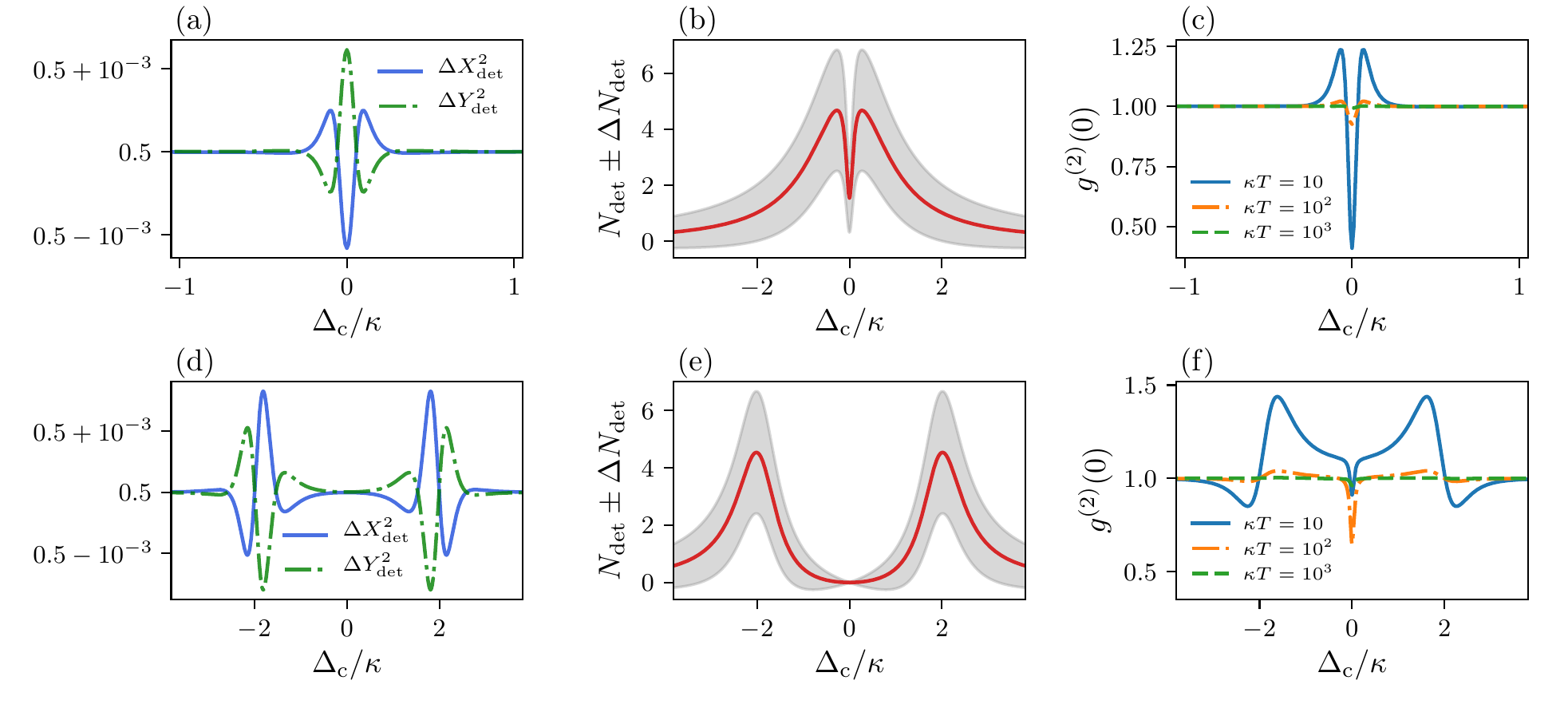}
\caption{\emph{Detected signal with a single emitter}. Plots of quadrature squeezing [(a),(d)], average photon number and variance (shaded area) [(b),(e)], and the photon correlation function $g^{(2)}(0)$ [(c),(f)] for a laser scan around the resonance (a)--(c) in the Purcell regime ($g=0.2\kappa$) and (d)--(f) in the strong-coupling regime ($g=2\kappa$), respectively. The remaining parameters here are $\gamma = \kappa/20$, $\eta=\kappa/20$ and $T=10^3\kappa^{-1}$. In panels (c) and (f), the effect of increasing integration time on bunching and antibunching is illustrated.}
\label{fig3}
\end{figure*}

\subsection{Time-integrated signal detection}
We define detected field operators at some time $t$ (chosen already after reaching steady state) by integrating over the continuous output fields during the detection window $t-T$ to $t+T$. While in Ref.~\cite{plankensteiner2017cavity} we have analyzed the antiresonance behavior in terms of the classical average of the intracavity field amplitude, we can here compute expectation values of the detected field photon number operator and its variance. We provide analytical expressions for these quantities as well as for the quadrature variances and the second-order correlation function at zero time $g^{(2)}(0)$.

\subsubsection{Classical signal}
The classical detected signal is defined as the time integral (over the detection window $2T$) of the continuous output field amplitude expectation value. The reflected signal is
\begin{align}
\langle \A\ts{det}(t) \rangle =\frac{1}{\sqrt{2T}}\int_{t - T}^{t + T} dt' ~\langle \A\ts{out}(t')\rangle = \sqrt{\frac{2T\eta^2}{\kappa}}r\ts{c},
\end{align}
while in transmission one detects
\begin{align}
\langle \B\ts{det}(t) \rangle =\frac{1}{\sqrt{2T}}\int_{t - T}^{t + T} dt' ~\langle \B\ts{out}(t')\rangle = \sqrt{\frac{2T\eta^2}{\kappa}}t\ts{c}.
\end{align}
The definition of the operators above fulfills the canonical commutation relations $[\A\ts{det}, \A\ts{det}^\dag]=1$ and $[\B\ts{det}, \B\ts{det}^\dag]=1$. The transmission of the cavity shows the signature of the antiresonance (both in amplitude as a dip and in phase as a rapid phase switch when the laser is swept across the common emitter-field resonance) as it is simply proportional to the cavity transmission function derived in Ref.~\cite{plankensteiner2017cavity}. Notice that for weak pumping, especially around the resonance dip, the integration time has to be large in order to distinguish the classical signal from shot noise.

\subsubsection{Fluctuation correlation matrix of the detected field}
According to the approach we employed to obtain higher order correlations of the output, let us define a vector of detected zero-average operators,
\begin{align}
\textbf{v}\ts{det}(t) &= \frac{1}{\sqrt{2T}}\int_{t - T}^{t + T} dt' ~\textbf{v}\ts{out}(t').
\end{align}
Note, that the component $v^1\ts{det}(t)=\hat{a}\ts{det}(t)$ is the detected signal fluctuation operator in reflection, and similarly $v^3 \ts{det}(t)=\hat{b}\ts{det}(t)$ is the detected signal fluctuation operator in transmission. Two-point correlations are needed in order to find the expectation value of the photon number. Therefore, one has to relate the correlation matrix of the output fields with the intra-cavity correlations. As a general formulation, we write the whole correlation matrix of the detected quantities as
\begin{align}
\textbf{V}\ts{det}(t) &= \frac{1}{2T}\int_{t - T}^{t + T}dt'\int_{t - T}^{t + T}dt''\braket{\textbf{v}\ts{out}(t')\textbf{v}\ts{out}(t'')^\top}.
\end{align}
We now use the Fourier transformation of the output operators (see Appendix \ref{sec:App.Fourier}) to relate the detected correlation matrix to the spectrum matrix of output operators,
\begin{align}
\label{Vdet1}
\textbf{V}\ts{det} &= \frac{1}{\pi T}\int_{-\infty}^{\infty}d\omega \left[\frac {\sin^2{\omega T}}{\omega^2}\right]\pmb{\cal{S}}\ts{out}(\omega).
\end{align}
In general, one can already compute the correlation matrix from this expression. However, for long integration times (longer than the inverse of the characteristic linewidth of the spectrum), the sinc-function inside the integral picks out only frequencies close to zero (around the laser frequency). This allows one to replace the sinc function with a $\delta$ function and the detected correlation matrix is given by the simple expression
\begin{align}
\label{Vdet2}
\textbf{V}\ts{det} &= \pmb{\cal{S}}\ts{out}(0).
\end{align}
We will use this result in the following subsections to derive expressions in terms of matrix elements of $\pmb{\cal{S}}\ts{out}(0)$ for the variance in quadratures, photon number expectation value and variance, and the second-order correlation function. The detected time-integrated quadratures in transmission are
\begin{subequations}
\begin{align}
\hat{X}\ts{det} &= \frac{1}{\sqrt{2}}\left(\B\ts{det} + \B\ts{det}^\dag\right),
\\
\hat{Y}\ts{det}&= \frac{-i}{\sqrt{2}}\left(\B\ts{det} - \B\ts{det}^\dag\right),
\end{align}
\end{subequations}
with similar expressions for the reflected field. With the help of the above expressions for the correlation matrix, one can compute their respective variances as
\begin{subequations}
\begin{align}
\Delta X\ts{det}^2 =\frac{1}{2} +{\cal{S}}^{43} \ts{out}(0) + \Re \left[ {\cal{S}}^{33}\ts{out}(0)\right],\\
\Delta Y\ts{det}^2 =\frac{1}{2} +{\cal{S}}^{43} \ts{out}(0) - \Re \left[ {\cal{S}}^{33}\ts{out}(0)\right].
\end{align}
\end{subequations}
In \fref{fig3}(a) the detected quadrature variances are shown around the cavity resonance (by scanning the laser frequency around it) exhibiting small squeezing properties around the antiresonance dip. While the squeezing in the strong coupling regime depicted in \fref{fig3}(d) is approximately of the same magnitude, it is shifted by $\pm g$ from the cavity and emitter resonance, i.e., the squeezing occurs at the polariton resonances.

\subsubsection{Photon number and its variance.}
The detected photon number operator for the transmission is
\begin{align}
\hat{N}\ts{det}(t) &= \frac{1}{2T}\int_{t - T}^{t + T} dt' \int_{t - T}^{t + T} dt'' ~\B^\dag\ts{out}(t')\B\ts{out}(t''),
\end{align}
with an expectation value
\begin{align}
\label{N_det_avg}
N\ts{det} = \langle \hat{N}\ts{det}(t) \rangle = |\langle \B\ts{det}(t) \rangle|^2 + \langle \hat{b}^\dagger\ts{det}\hat{b}\ts{det}\rangle.
\end{align}
Notice that in the absence of any nonlinear terms in the evolution, the detected photon number would be simply given by the absolute square of the classical amplitude as is characteristic for coherent states. However, the second term in Eq. \eqref{N_det_avg} is nonzero and can again be expressed in terms of the output as $\langle \hat{b}^\dagger\ts{det}\hat{b}\ts{det} \rangle ={\cal{S}}^{43}\ts{out} (0)$. We can then analyze the behavior of the photon number at the detector which is plotted for variable laser drive frequency (around the antiresonance) in Fig.~\ref{fig3}(b) and for the strong coupling regime in Fig.~\ref{fig3}(e). The behavior, as expected, mimics the cavity transmission profile. More interesting aspects emerge when one analyzes the variance of the photon number around the average; to this end we explicitly write the expression for the variance,
\begin{align}
&\left[\Delta N\ts{det}(t)\right]^2=\langle \hat{b}^\dagger\ts{det}\hat{b}\ts{det}\hat{b}^\dagger\ts{det}\hat{b}\ts{det} \rangle-{\langle \hat{b}^\dagger\ts{det}\hat{b}\ts{det} \rangle}^2 +
\notag \\
\label{variance}
&+|\langle \B\ts{det}\rangle|^2\left[1+2{\langle \hat{b}^\dagger\ts{det}\hat{b}\ts{det} \rangle}\right]+{{\langle \B\ts{det} \rangle}^*}^2 \langle \hat{b}^2\ts{det} \rangle +
\\
&+{\langle \B\ts{det}\rangle}^2\langle (\hat{b}\ts{det}^\dagger)^2 \rangle.
\notag
\end{align}
The two-operator averages emerge immediately from the spectrum matrix as $\langle \hat{b}^2 \ts{det} \rangle = {\cal{S}}^{33}\ts{out} (0)$ and $\langle (\hat{b}\ts{det}^\dagger)^2\rangle = {\cal{S}}^{44}\ts{out} (0)$. The task of evaluating four-point correlations is a bit more cumbersome. However, we can apply Isserlis' theorem to the output (see Appendix \ref{sec:App.Fourier}) to express any four-operator correlations as sums over all permutations of two operator correlations. After time integration one derives an according expansion for four detected operator correlations. This allows one to compute quantities such as
\begin{align}
\langle \hat{b}^\dagger\ts{det}\hat{b}\ts{det}\hat{b}^\dagger\ts{det}\hat{b}\ts{det} \rangle &= {\cal{S}}^{43}\ts{out}(0){\cal{S}}^{43}\ts{out}(0)+{\cal{S}}^{44}\ts{out}(0){\cal{S}}^{33}\ts{out}(0)
+
\notag \\
&+ {\cal{S}}^{43}\ts{out}(0){\cal{S}}^{34}\ts{out}(0).
\end{align}
Finally, the expression for the detected photon number variance after replacement of two- and four-operator correlations in Eq.~\eqref{variance} becomes
\begin{align}
&\left[\Delta {N}\ts{det}(t)\right]^2 = \left|{\cal{S}}^{44} \ts{out}(0) \right|^2 + |\langle \B\ts{det} \rangle|^2 [1+2{\cal{S}}^{43}\ts{out}(0)]+
\notag \\
&+2 \Re \left\{  {\langle \B\ts{det}\rangle}^2  {\cal{S}}^{44}\ts{out}(0)\right\} + {\cal{S}}^{43}\ts{out}(0){\cal{S}}^{34}\ts{out}(0).
\end{align}
The result is illustrated in~\fref{fig3}(b) (in the antiresonance regime) and in~\fref{fig3}(e) (in the strong coupling regime). The variance is included as a shaded region around the mean photon number. Owing to the weak coupling and large integration time, the result corresponds to the standard shot noise of a detected coherent state. The contribution from the two-photon terms ${\cal{S}}^{33}$ and ${\cal{S}}^{44}$, while showing up as quadrature squeezing, have little effect on the photon number variance.

\subsubsection{Second-order correlation function of the photon number}
In order to understand the photon statistics of the transmitted light from the cavity, we calculate the second-order correlation function~\cite{xu2017intensity} at zero delay $g^{(2)}(0)$ defined by
\begin{align}
g^{(2)}(0) &= \frac{\langle \B\ts{det}^\dag \B\ts{det}^\dag \B\ts{det} \B\ts{det}\rangle}{\langle \B\ts{det}^\dag \B\ts{det}\rangle^2}.
\end{align}
Note, that for a coherent state $g^{(2)}(0)=1$, which is characteristic for Poissonian light. However, terms in the output spectrum such as ${\cal{S}}^{44}\ts{out}(0)$ denote the presence of photon-photon correlations coming from non-vanishing expectation values of $\langle \hat{b}^\dagger\ts{det}\hat{b}^\dagger\ts{det}\rangle$. After evaluating the different terms in the expression above and rewriting all occurring four-point correlations as before (see Appendix \ref{sec:App.Fourier}), one finds
\begin{widetext}
\begin{align}
g^{(2)}(0) = \frac{|\langle \B\ts{det} \rangle|^4 + 4|\langle \B\ts{det} \rangle|^2{\cal{S}}^{43}\ts{out}(0) + 2\Re\left\{ \langle \B\ts{det} \rangle^2{\cal{S}}^{44}\ts{out}(0)\right\} + |{\cal{S}}^{44}\ts{out}(0)|^2 + 2{\cal{S}}^{43}\ts{out}(0){\cal{S}}^{43}\ts{out}(0)}{|\langle \B\ts{det} \rangle|^4 + 2|\langle \B\ts{det} \rangle|^2{\cal{S}}^{43}\ts{out}(0) + {\cal{S}}^{43}\ts{out}(0){\cal{S}}^{43}\ts{out}(0)}.
\end{align}
\end{widetext}
For different detection time windows, the behavior of the second-order correlation function is shown in Fig.~\ref{fig3}(c) (for the antiresonance regime) and in Fig.~\ref{fig3}(f) (for the strong coupling regime). Longer detection times have the tendency of washing out the photon bunching and antibunching effects.

\subsection{Nonlinear effects}

\begin{figure*}[t]
\includegraphics[width=\textwidth]{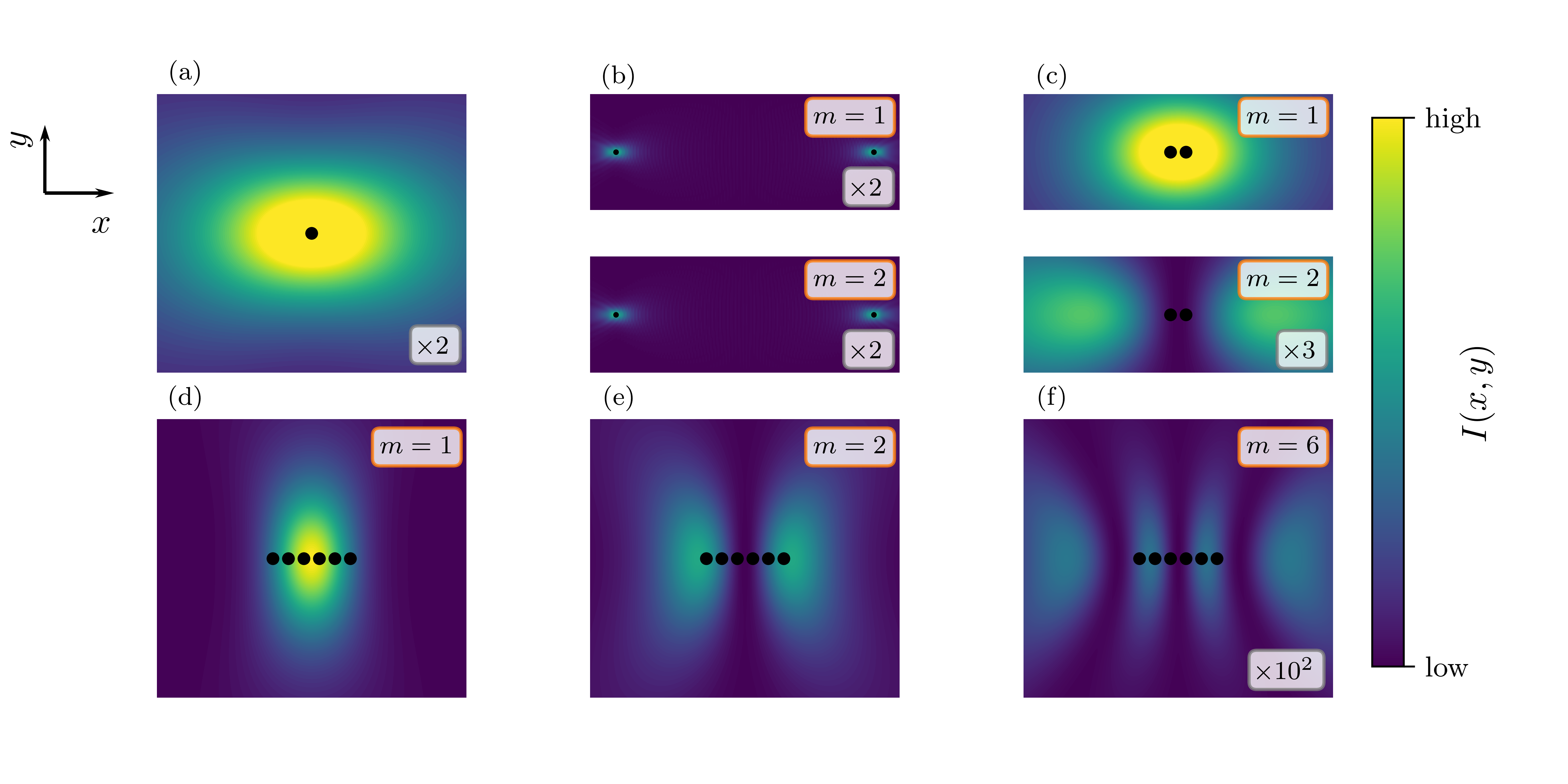}
\caption{\emph{Free space radiation patterns of exciton states}. The spatial intensity profile of the emitted field is plotted in arbitrary units, but to scale. (a) For a single emitter, we observe the standard dipole radiation. (b) For two emitters placed far apart ($d\gg\lambda\ts{e}$), they hardly interact and the chosen state therefore has little effect on the emitted field. (c) When the two emitters are placed much closer ($d=0.3\lambda\ts{e}$) than a single wavelength, the interactions become very strong, leading to superradiant loss ($m=1$) and subradiance ($m=2$), where the emitted field is predominantly radiated into the axis along which the emitters are placed. (d)--(f) The same effect is even more distinct for a closely spaced ($d=0.3\lambda\ts{e}$) equidistant chain of emitters (here $N=6$). With increasing phase asymmetry (increasing $m$) there are more field nodes. Note that some of the shown field intensities had to be scaled since they are orders of magnitude smaller than the superradiant field of the chain. This is indicated (where needed) by the scaling factor in the bottom right. The dipole moments have been chosen along the $y$ axis and the profile in the $x$-$y$ plane is observed at a transverse distance $z=2\lambda\ts{e}$.}
\label{fig4}
\end{figure*}

A single quantum emitter is a nonlinear object as its response (the amplitude of the stimulated transition dipole moment) is not only proportional to the driving field amplitude. In the next order of approximation, a small component emerges from the AC Stark shift of the excited state level proportional to the field intensity, the so-called Kerr effect. At the macroscopic level, this effect is seen as a modification of the index of refraction with increasing light field intensities. For the hybrid cavity-emitter system, we analyze the response of the transition dipole moment to the driving laser amplitude $\eta$. As opposed to the bare free space nonlinearity expected from a two-level system, the cavity can lead to a modified "vacuum-dressed emitter" nonlinearity. Inside the cavity, we analytically derive this small correction by assuming that $z=(2|\beta|^2-1)$ and obtaining the new steady-state solution from
\begin{subequations}
\begin{align}
0 &= -(\kappa-i\Delta_{\mathrm{c}}) \alpha +\eta-ig\beta,\\
0 &= -(\gamma-i\Delta_{\mathrm{e}})\beta + i g \alpha (2|\beta|^2-1).
\end{align}
\end{subequations}
We can find a solution for $\beta=\beta^{(1)}+\beta^{(3)}$ where the linear term $\beta^{(1)}$ is the previously derived response of the emitter's dipole in Eq.~\eqref{dipolelinear} proportional to $\eta$. The next order correction is

\begin{equation}
\beta^{(3)} = -2\beta^{(1)}|\beta^{(1)}|^2\left(1-\frac{ig}{\eta}\beta^{(1)}\right).
\end{equation}
The Kerr nonlinearity is proportional to the field intensity $\eta^2$ (we considered the field amplitude real) and leads to a modification of the cavity transmission function from the computed $t^{(1)}$ expression in Eq.~\eqref{trans},

\begin{align}
t\ts{c} = & t\ts{c}^{(1)}\left(1 + \frac{2g^2\left|\beta^{(1)} \right|^{2}\left[1-i(g/\eta)\beta^{(1)}\right]}{(\gamma-i\Delta_{\mathrm{e}})(\kappa-i\Delta_{\mathrm{c}})}\right).
\end{align}
Notice that the nonlinearity matches the behavior of the linear response in that it is largest around the antiresonance. Maximal nonlinear response occurs when the linear one is maximal as well. We can then find a simple and instructive expression $\beta^{(1)}(\Delta=0)=-iC/(1+C) \eta/g$, which shows that an increase in the cooperativity (by, for example, suppressing the radiative rate while keeping $g$ constant) brings the nonlinearity to a saturation value.

\section{Free space collective dynamics: super- and subradiant states}
\label{freespace}

Before analyzing the physics of a cavity mode interacting with an ensemble of coupled quantum emitters, let us briefly review some properties of the bare coupled emitter ensemble (in free space). In general, it is not possible to diagonalize the Hamiltonian including dipole-dipole interactions. However, a common approach is to truncate the Hilbert space at small or even single excitations \cite{plankensteiner2015selective,zoubi2008bright}. Then, at extremely small distances, one can use the fact that $|\Omega_{ii+1}|\gg|\Omega_{ii+2}|$ to make the nearest-neighbor (NN) approximation. The full Hamiltonian then becomes a tridiagonal symmetric T\"oplitz matrix, which can be analytically diagonalized. The resulting set of eigenstates $\{\ket{m}\}_{m=1}^N$ is given by
\begin{align}
\ket{m} = \sqrt{\frac{2}{N + 1}}\sum_j\sin\left(\frac{\pi m j}{N+1}\right)\hat{\sigma}_j^+ \ket{g}^{\otimes N}.
\end{align}
They correspond to collective excitations of different symmetries with corresponding energies
\begin{align}
\label{m_energies}
\omega_m &= \omega\ts{e} + 2\Omega_{12}\cos\left(\frac{\pi m}{N +1 }\right),
\end{align}
ranging from $- 2\Omega_{12}$ to $2\Omega_{12}$ around the bare noninteracting energy $\omega\ts{e}$. These expressions illustrate what is required to faithfully excite a specific collective state: One needs to match both the local phases given by the coefficients of the states as well as the shifted resonance energies. While the latter is quite straightforward, addressing an ensemble of emitters with large local phase differences within a small volume can be challenging. A number of proposals on how this could be achieved have been brought forward in recent studies. The suggested schemes involve among others a magnetic field gradient~\cite{plankensteiner2015selective} or light that imprints the phases due to a polarization gradient~\cite{plankensteiner2017cavity,jen2018directional}.

The reason for these extensive studies of preparation schemes is that large phase differences cause the emitter dipole fields to interfere destructively [see \fref{fig4}(f)], thus yielding an extremely small total field. Thereby, the lifetime of states with large phase differences (subradiant states) is vastly enhanced, making them ideal candidates for precision spectroscopy or quantum memories. This gain in lifetime becomes even more significant when considering the fact that symmetric excitation in dense emitter ensembles leads to superradiance \cite{dicke1954coherence}, which in the limit of vanishing separation leads to a factor $N$ enhancement in spontaneous emission.

This behavior due to phase (a-)symmetry can be investigated by computing the field radiated by the dipoles of the emitters taking into account the fact that they interfere with one another. Namely, the free electric field is just
\begin{align}
\hat{\textbf{E}}(\textbf{r},t) = \hat{\textbf{E}}^{(+)}(\textbf{r},t) + \text{H.c.},
\end{align}
where
\begin{align}
\hat{\textbf{E}}^{(+)}(\textbf{r},t) &= \sum_{\textbf{k},\lambda} \sqrt{\frac{\hbar\omega_k}{2\epsilon_0 V}}\textbf{e}_{\textbf{k},\lambda} \hat{A}_{\textbf{k},\lambda}(t)e^{i\textbf{k}\cdot \textbf{r}}.
\end{align}
Because of the dipole coupling of all emitters to the field, the photon annihilation operators simply follow the emitter coherence operators,
\begin{align}
\dot{\hat{A}}_{\textbf{k},\lambda}(t) &= -ig_{\textbf{k},\lambda}\sum_j\hat{S}_j(t)e^{-i\textbf{k}\cdot \textbf{r}_j} e^{i(\omega_k - \omega\ts{e})t}.
\end{align}
Resolving the sum over wave vectors as an integral (due to the density of modes) and in addition making the Markov approximation allows one to find an expression for the electric field containing only emitter operators and geometric factors (see Appendix \ref{E-field-derivation} for details).

\begin{figure}[t]
\includegraphics[width=0.95\columnwidth]{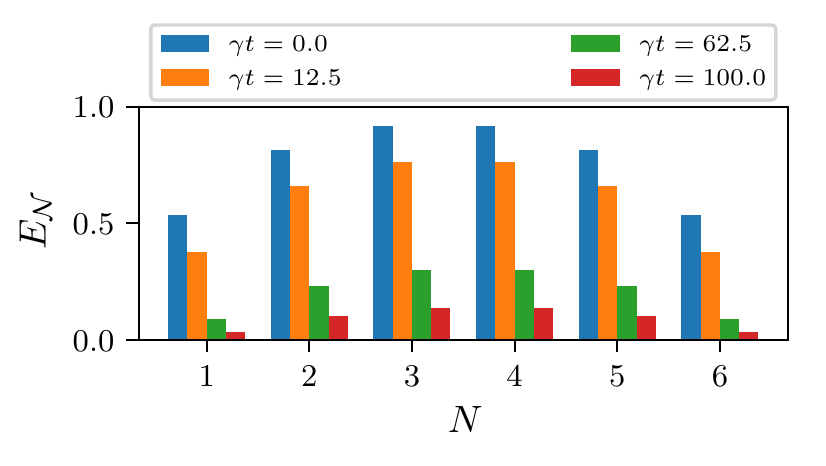}
\caption{\emph{Entanglement of emitters in a free space chain in a subradiant state}. A six-emitter equidistant chain with a separation of $d=0.1\lambda\ts{e}$ is initialized in the state $\ket{m=N}$ and left to evolve freely. We plot the logarithmic negativity of each emitter with respect to the $N-1$ remaining emitters in the chain at different point in times (the left most bar for each site corresponds to $\gamma t=0$).}
\label{fig5}
\end{figure}

The intensity of the resulting field is illustrated in \fref{fig4} for a single emitter, two emitters and a chain of emitters. While for a single emitter we observe dipole radiation [see \fref{fig4}(a)], interference occurs when more than one emitter is present. If the separation between emitters is large, there are no substantial interactions enhancing or suppressing radiation. As illustrated in \fref{fig4}(b), the radiated field is then independent of the chosen state. As soon as the emitters are close (separation smaller than half a wavelength), the state in which they are prepared has a significant effect on the field. The radiated intensity is either enhanced due to constructive interference (superradiance) or suppressed due to destructive interference (subradiance).

These phenomena are even more dominant for more emitters. In \fref{fig4}(d)--(f), we see the radiated intensity for different choices of $m$ for an equidistant chain of $N=6$ emitters. While the field is quite large in the symmetric case ($m=1$), we can see that for other choices of $m$, destructive interference occurs. Namely, for the smallest asymmetric choice of states, $m=2$, there is only one change in sign of the phase which occurs in the middle of the chain. It is clear that at this point the fields radiated by each half of the chain cancel each other. For $m=N$ this effect culminates in maximal destructive interference, which widely inhibits spontaneous emission from the chain.

Another property of these collective states (both super- and subradiant ones) is that they commonly feature high degrees of entanglement \cite{plankensteiner2015selective, hebenstreit2017subradiance}. As such, they form an interesting resource for quantum information processing, where highly subradiant states are even more useful due to the increased lifetime of correlations.

Even though subradiant systems show only moderate two-pair correlations, the overall entanglement is large. Specifically, each of the emitters is highly entangled with all the other emitters. In order to illustrate this point, we plot the logarithmic negativity \cite{plenio2005logarithmic}, which is an entanglement monotone. For a bipartite system consisting of the subsystems $A$ and $B$, it is defined as
\begin{align}
E_{\mathcal{N}}(\rho) &= \log_2\left(|\rho^{T_A}|\right),
\end{align}
where $\rho^{T_A}$ denotes the partial transpose with respect to the subsystem $A$ and $|\cdot|$ is the tracenorm. In \fref{fig5}, we initialize a chain of emitters in the state with the highest phase difference ($m=N$) and let it evolve freely over time. At distinct time points, we compute the logarithmic negativity for each emitter (i.e., we choose our bipartite system to consist of the $i$th emitter and the rest of the chain). One can see that the amount of entanglement is even in the initial state significantly larger in the center of the chain. Over time, this behavior is retained, and correlation is only slowly lost due to excitation loss of the chain. Even at $t=100\gamma^{-1}$, there still is considerable entanglement in the system.

\section{Spectroscopy of the collective Purcell effect}
\label{collective}

We now generalize the formalism developed for the single quantum emitter case to many, coupled quantum emitters with special focus on addressing collective subradiant states. In a first step, we derive the cavity transmission function in the linear regime showing the occurrence of collective resonances of different radiative natures (subradiant/superradiant) and the scaling of the cooperativity when proper illumination techniques (matching phase and energy of the collective subradiant resonances) are employed. We then look at collective cooperative effects on output field squeezing, photon-photon correlations, and the enhancement of the overall ensemble Kerr nonlinearity. We find that in all these investigations, enhancement is always reached in the \textit{cooperative collective} regime (where the interacting ensemble shows a much higher cooperativity than a noninteracting ensemble).

\subsection{Subradiant enhancement of cavity-emitter cooperativity}

\begin{figure}[t]
\includegraphics[width=\columnwidth]{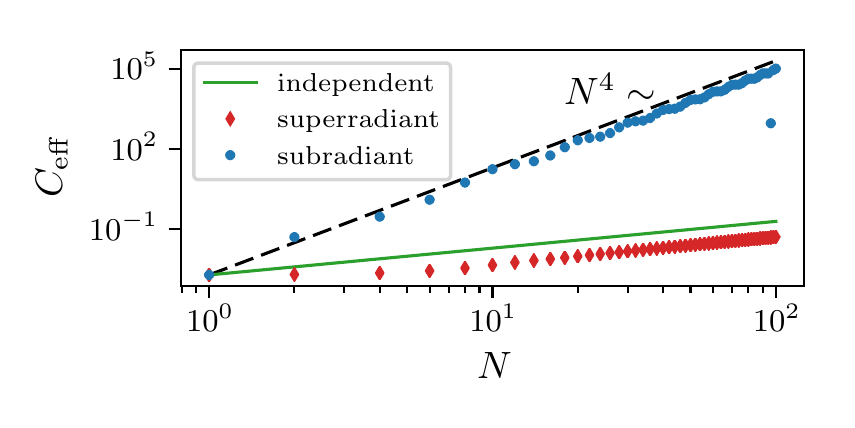}
\caption{\emph{Scaling of the effective cooperativity with the number of emitters in an equidistant chain}. Strong collective effects are present when the emitters are closely spaced at $d=0.1\lambda\ts{e}$, leading to distinct scalings of the cooperativity. As a reference, we plot the cooperativity for independent emitters, which scales linearly with the number of emitters. The dashed line indicates the scaling of the subradiant case with $N^4$ (note that a simple polynomial fit returns a scaling of approximately $N^{3.81\pm 0.01}$). The super- and subradiance is caused by symmetric ($g_i=g$) and asymmetric [$g_i=(-1)^ig$] coupling, respectively. We chose the parameters $\gamma=\kappa/20=5g$ such that for all $N$ we are in the regime where $NC\ll 1$.}
\label{fig6}
\end{figure}

In order to perform a classical analysis of the response of a cavity weakly coupled to $N$ interacting quantum emitters (deriving the amplitude transmission), it suffices to solve the coupled equations of motion for classical averages. In a compact matrix form, this is written as the following equations of motions \cite{plankensteiner2017cavity},
\begin{subequations}
\begin{align}
\dot \alpha &= -\left(\kappa - i\Delta_\text{c}\right)\alpha + \eta - i\textbf{G}\cdot\boldsymbol{\beta},
\\
\dot{\boldsymbol{\beta}} &= i \Delta\ts{e} \boldsymbol{\beta} - i \boldsymbol{\Omega}\boldsymbol{\beta} - i \textbf{G}\alpha - \boldsymbol{\Gamma}\boldsymbol{\beta},
\end{align}
\end{subequations}
where now $\boldsymbol{\beta}$ and $\textbf{G}$ are column vectors with entries $\beta_i$ and $g_i$. The matrices $\boldsymbol{\Omega}$ and $\boldsymbol{\Gamma}$ have the elements $\Omega_{ij}$ and $\gamma_{ij}$, respectively. In steady state, the transmission coefficient for the cavity amplitude reads
\begin{align}
t\ts{c}  &= \frac{\kappa}{-i\Delta\ts{c}+\kappa + \textbf{G}^\top\textbf{G}/[-i\Delta_\text{eff}(\Delta\ts{e}) + \gamma_\text{eff}(\Delta\ts{e})]},
\end{align}
where the effective $\Delta\ts{e}$-dependent collective energy shifts and linewidths are given by
\begin{subequations}
\begin{align}
\label{delta_eff}
\Delta_\text{eff}(\Delta\ts{e}) &= \Im\left\{\frac{\textbf{G}^\top\textbf{G}}{\textbf{G}^\top\left(-i\Delta\ts{e}\mathbbm{1}+i\boldsymbol{\Omega}+\boldsymbol{\Gamma}\right)^{-1}\textbf{G}}\right\},
\\
\gamma_\text{eff}(\Delta\ts{e}) &= \Re\left\{\frac{\textbf{G}^\top\textbf{G}}{\textbf{G}^\top \left(-i\Delta\ts{e}\mathbbm{1}+i\boldsymbol{\Omega}+\boldsymbol{\Gamma}\right)^{-1}\textbf{G}}\right\}.
\label{gamma_eff}
\end{align}
\end{subequations}
In analogy to the single emitter case we can define an effective $N$-emitter cooperativity by
\begin{align}
\label{CN_def}
C\ts{eff}(\Delta\ts{e}) = \frac{\textbf{G}^\top\textbf{G}}{\kappa\gamma_\text{eff}(\Delta\ts{e})}.
\end{align}
The message of the above equation is that the numerator and denominator no longer share the same dependency on the individual emitter properties (such as the dipole moment). Thus, a much larger effective cooperativity can be reached if one manages to excite a subradiant collective state for which the effective decay rate is small.

Note that, as mentioned before, in addition to matching the symmetry of the collective state one wants to address, one also has to match the state's energy. The cavity has to be tuned to fulfill the condition $\Delta\ts{eff}(\omega\ts{c}-\omega\ts{e})=0$, such that at the point of resonance where $\omega\ts{c}=\omega_{\ell}$ the collective state is also resonant. This is straightforward to do numerically. The distinct scaling of the cooperativity is shown in \fref{fig6}. It can be seen that the subradiant enhancement of the effective cooperativity shows a beneficial scaling with the number of emitters with approximately $N^4$. It has been shown that the lowest decay rate theoretically possible reduces exponentially with the number of emitters \cite{ostermann2014protected}. The most robust states that can be reached in reality, though, scale with $N^{-3}$ \cite{asenjo2017exponential}. This, combined with the collective enhancement of the coupling to the cavity mode with $N$, yields the scaling observed in \fref{fig6}. Deviations from the $N^4$ scaling are due to imperfect addressing of subradiant states as well as to finite-size effects.

On the other hand, the superradiant decay almost compensates the enhancement of the coupling with $N$ for small numbers of emitters, since in that case also the decay rate scales approximately linear. This keeps the effective cooperativity constant at the value for a single emitter. Again, due to imperfect resonance matching and finite-size effects, the decay rate does not show perfect linear scaling and saturates at some point. The enhanced coupling is therefore no longer perfectly compensated for larger $N$ and we again observe a linear enhancement of the cooperativity. However, the effective cooperativity affected by superradiance can never surpass the cooperativity of the same number of independent emitters.

Besides the number of quantum emitters, another parameter on which the collective modification of the cooperativity strongly depends is the distance between the emitters. It governs the strength of the dipole-dipole interactions and subsequently any enhancement or reduction of the light-emitter interactions. Systematic investigations of the dependence of super- and subradiance on the emitter separation have been performed \cite{asenjo2017exponential, ostermann2014protected}. At sufficiently small distances -- which is the regime we consider here -- the collective decay has been shown to be a monotonous function of the particle-particle separation \cite{ostermann2014protected}; i.e. collective effects increase as the distance between emitters decreases. Furthermore, disorder in the emitter positions only marginally affects subradiance and can even lead to more long-lived states \cite{plankensteiner2015selective}.

\subsection{Nonclassical collective effects in detected fields}
\label{sec:non-classical}

\begin{figure}[t]
\includegraphics[width=\columnwidth]{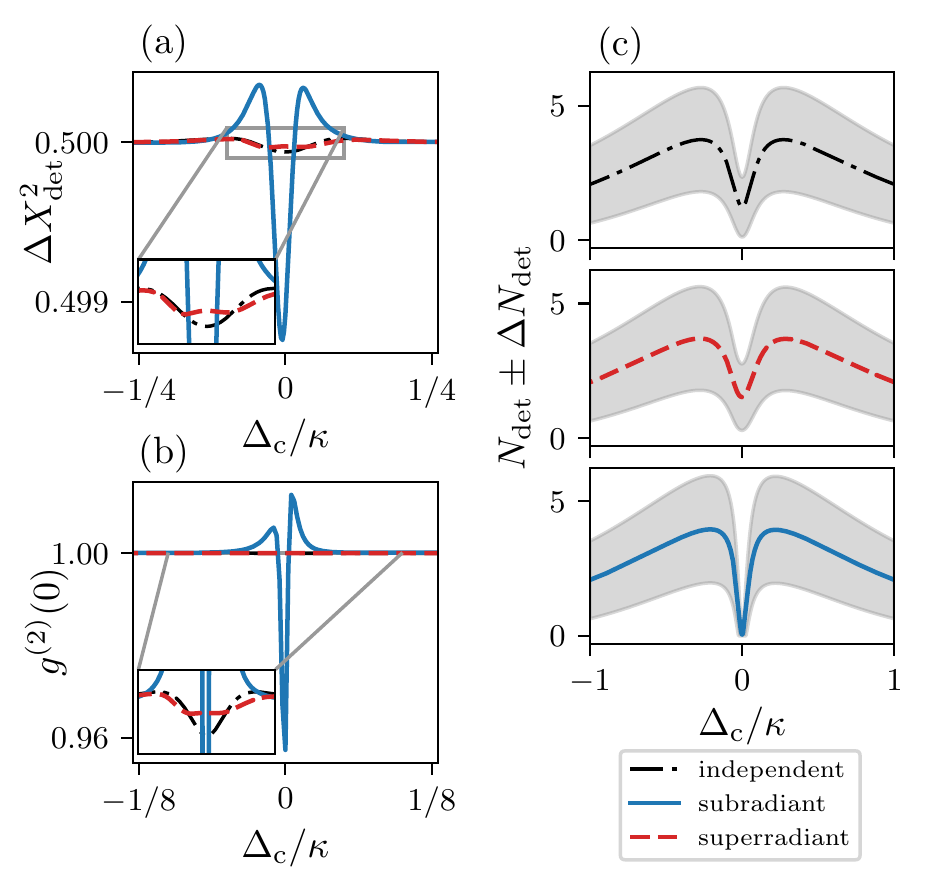}
\caption{\emph{Cooperative effects of four coupled emitters}. Plots of (a) the variance of the detected  $x$ quadrature, (b) the $g^{(2)}(0)$ function, and (c) the photon number with its variance for four coupled emitters around the antiresonance for subradiant, superradiant and independent cases. The parameters are $g=2\gamma=\kappa/10$, $\eta=\kappa/100$, and $T=2\times 10^{4}\kappa^{-1}$, with the emitter dipoles oriented perpendicular to the cavity axis and a separation of the emitters of $0.3\lambda\ts{e}$.}
\label{fig7}
\end{figure}

We can follow the same procedure as for a single emitter to investigate the squeezing at the output port. From Eq.~\eqref{QLE_single_vec}, we obtain the vector equations for the quantum fluctuations in the form
\begin{subequations}
\label{N-QLE-vec}
\begin{align}
\dot{\hat a} &= -\left(\kappa - i\Delta\ts{c}\right)\hat a - i\textbf{G}\cdot \boldsymbol{\hat{\sigma}} + \sqrt{\kappa}\hat a\ts{in} + \sqrt{\kappa}\hat b\ts{in},
\\
{\boldsymbol{\dot{\hat{\sigma}}}} &= \textbf{A}\boldsymbol{\hat{\sigma}} + \textbf{B}\boldsymbol{\hat{\sigma}}^z + i\textbf{G}_z\hat a - \boldsymbol{\bar{\xi}},
\\
\boldsymbol{\dot{\hat{\sigma}}}^z &= -2\gamma\boldsymbol{\hat{\sigma}}^z + \textbf{K}\boldsymbol{\hat{\sigma}} + \textbf{K}^{*}\boldsymbol{\hat{\sigma}}^\dag + 2i\textbf{G}_\beta\hat a^\dag - 2i\textbf{G}_\beta^* \hat a
\\
&+ 2\boldsymbol{\bar{\xi}}^z.
\notag
\end{align}
\end{subequations}
We have defined the modified coupling vectors $\textbf{G}_z = (z_1g_1,...,z_Ng_N)^\top$ and $\textbf{G}_\beta = (\beta_1g_1,...,\beta_Ng_N)^\top$. The coupling matrices are given by
\begin{subequations}
\begin{align}
\textbf{A}_{jk} &= -\left(\gamma - i\Delta\ts{e}\right)\delta_{jk} +
\\
& +\left(1-\delta_{jk}\right)\left(i\Omega_{jk} + \gamma_{jk}\right)z_j,
\notag \\
\textbf{B}_{jk} &= \delta_{jk}\big(ig_j\alpha + \sum_{l\neq j}\left(i\Omega_{jl} + \gamma_{jl}\right)\beta_l\big),
\\
\textbf{K}_{jk} &= \big(2ig_j\alpha^* - 2\sum_{l\neq j}\gamma_{jk}\beta_l^*\big)\delta_{jk} +
\\
&- 2\left(1-\delta_{jk}\right)\gamma_{jk}\beta_j^*.
\notag
\end{align}
\end{subequations}
We can now express Eqs.~\eqref{N-QLE-vec} in vector form, $\dot{\textbf{v}}=\textbf{M}\textbf{v} + \textbf{N}\textbf{v}\ts{in}$, in analogy with the single-emitter case, with the proper matrix definitions. The drift matrix is given by
\begin{align}
\textbf{M} &= \begin{pmatrix}
-\left(\kappa - i\Delta\ts{c}\right) & 0 & -i\textbf{G}^\top & \textbf{0}^\top & \textbf{0}^\top \\
0 & -\kappa - i\Delta\ts{c} & \textbf{0}^\top & i\textbf{G}^\top & \textbf{0}^\top \\
i\textbf{G}_z & \textbf{0} & \textbf{A} & \underline{\textbf{0}} & \textbf{B} \\
\textbf{0} & -i\textbf{G}_z & \underline{\textbf{0}} & \textbf{A}^* & \textbf{B}^* \\
-2i\textbf{G}_\beta^* & 2i\textbf{G}_\beta & \textbf{K} & \textbf{K}^* & -2\gamma\mathbbm{1}
\end{pmatrix},
\end{align}
where $\textbf{0}$ is a vector containing $N$ zeros and $\underline{\textbf{0}}$ is a $N\times N$ matrix with only zero elements. The matrix multiplying the input noise operators is
\begin{align}
\textbf{N} = \begin{pmatrix}
\sqrt{\kappa} & 0 & \sqrt{\kappa} & 0 & \textbf{0}^\top & \textbf{0}^\top & \textbf{0}^\top\\
0 & \sqrt{\kappa} & 0 & \sqrt{\kappa} & \textbf{0}^\top & \textbf{0}^\top & \textbf{0}^\top\\
\textbf{0} & \textbf{0} & \textbf{0} & \textbf{0} & -\sqrt{2\gamma}\mathbbm{1} & \underline{\textbf{0}} & \underline{\textbf{0}}\\
\textbf{0} & \textbf{0} & \textbf{0} & \textbf{0} & \underline{\textbf{0}} & -\sqrt{2\gamma}\mathbbm{1} & \underline{\textbf{0}}\\
\textbf{0} & \textbf{0} & \textbf{0} & \textbf{0} & \underline{\textbf{0}} & \underline{\textbf{0}} & \sqrt{2\gamma}\mathbbm{1}
\end{pmatrix}.
\end{align}
Finally, the input noise correlation matrix in the many emitter case is
\begin{align}
\label{C_many-emitters}
\textbf{C} =
\begin{pmatrix}
0 & 1 & 0 & 0 & \textbf{0}^\top & \textbf{0}^\top & \textbf{0}^\top\\
0 & 0 & 0 & 0 & \textbf{0}^\top & \textbf{0}^\top & \textbf{0}^\top\\
0 & 0 & 0 & 1 & \textbf{0}^\top & \textbf{0}^\top & \textbf{0}^\top\\
0 & 0 & 0 & 0 & \textbf{0}^\top & \textbf{0}^\top & \textbf{0}^\top\\
\textbf{0} & \textbf{0} & \textbf{0} & \textbf{0} & \underline{\textbf{0}} & \textbf{C}_{\beta\beta} & \textbf{C}_{\beta z}\\
\textbf{0} & \textbf{0} & \textbf{0} & \textbf{0} & \underline{\textbf{0}} & \underline{\textbf{0}} & \underline{\textbf{0}}\\
\textbf{0} & \textbf{0} & \textbf{0} & \textbf{0} & \underline{\textbf{0}} & \textbf{C}_{\beta z}^\dag & \textbf{C}_{zz}
\end{pmatrix},
\end{align}
where $\textbf{C}_{\beta\beta}$ is the correlation matrix whose elements are given by Eq.~\eqref{noise-corr-1}, and analogous for the other indexed $\textbf{C}$ matrices (see Appendix~\ref{N-emitters-matrices}). Note that $\textbf{C}_{\beta z} = \textbf{C}_{z\beta}^\dag$.

With the help of these matrix definitions, it is straightforward to compute all figures of interest. All expressions for the field operators and variances are the same as for the single-emitter case; we merely have to insert the corresponding output spectrum matrix elements obtained for the many emitter case. The output spectrum matrix can be computed with the matrices above using Eq.~\eqref{S_out}.

The resulting higher order averages are shown in \fref{fig7}. There, we compare four independent, subradiant [asymmetric coupling, $g_i=(-1)^i g$] and superradiant (symmetric coupling, $g_i=g$) emitters. Note that, as in the discussion of the effective cooperativity, we match the cavity to the addressed collective state by choosing the cavity resonance such that $\Delta\ts{eff}(\omega\ts{c}-\omega\ts{e})=0$.

The detected photon number and its standard deviation is depicted in \fref{fig7}(c) and, as in the single-emitter case, it exhibits the same behavior as the transmission obtained from the average amplitude \cite{plankensteiner2017cavity}. The superradiance broadens and lessens the antiresonance, while the subradiant emitters lead to a very deep but narrow antiresonance.

As for the quadrature variance and second-order coherence function shown in \fsref{fig7}(a) and \ref{fig7}(b), the clear overall point here is this: Compared to independent emitters, superradiance (slightly) diminishes all nonlinear effects (see insets), while subradiance offers enhancements by orders of magnitude. Squeezing and antibunching both occur around the resonance (as for the single emitter in the weak coupling regime). However, both effects are distinctly larger in a frequency range much smaller than the cavity linewidth when the emitters are in a subradiant state. This enhancement is a signature of the counterintuitive effect that subradiant systems exhibit stronger classical dipoles: The stationary excited state population and thus the stationary magnitude of the induced dipole moment are inversely proportional to the spontaneous decay rate of the considered state. Therefore, as we see in \fref{fig7}, nonlinear effects such as squeezing and antibunching increase in consequence of the inhibited decay of a subradiant state. In contrast to this, superradiant states exhibit smaller stationary values for their collective dipole moments. However, the broadening of the effective linewidth due to superradiance is less prominent compared to the suppression of the decay in subradiant states. Going from the independent emitter to the superradiant case, there is thus a decrease in these quantum effects. Compared to the large enhancement due to subradiance, however, this reduction is somewhat less significant.

\begin{figure}[t]
\includegraphics[width=\columnwidth]{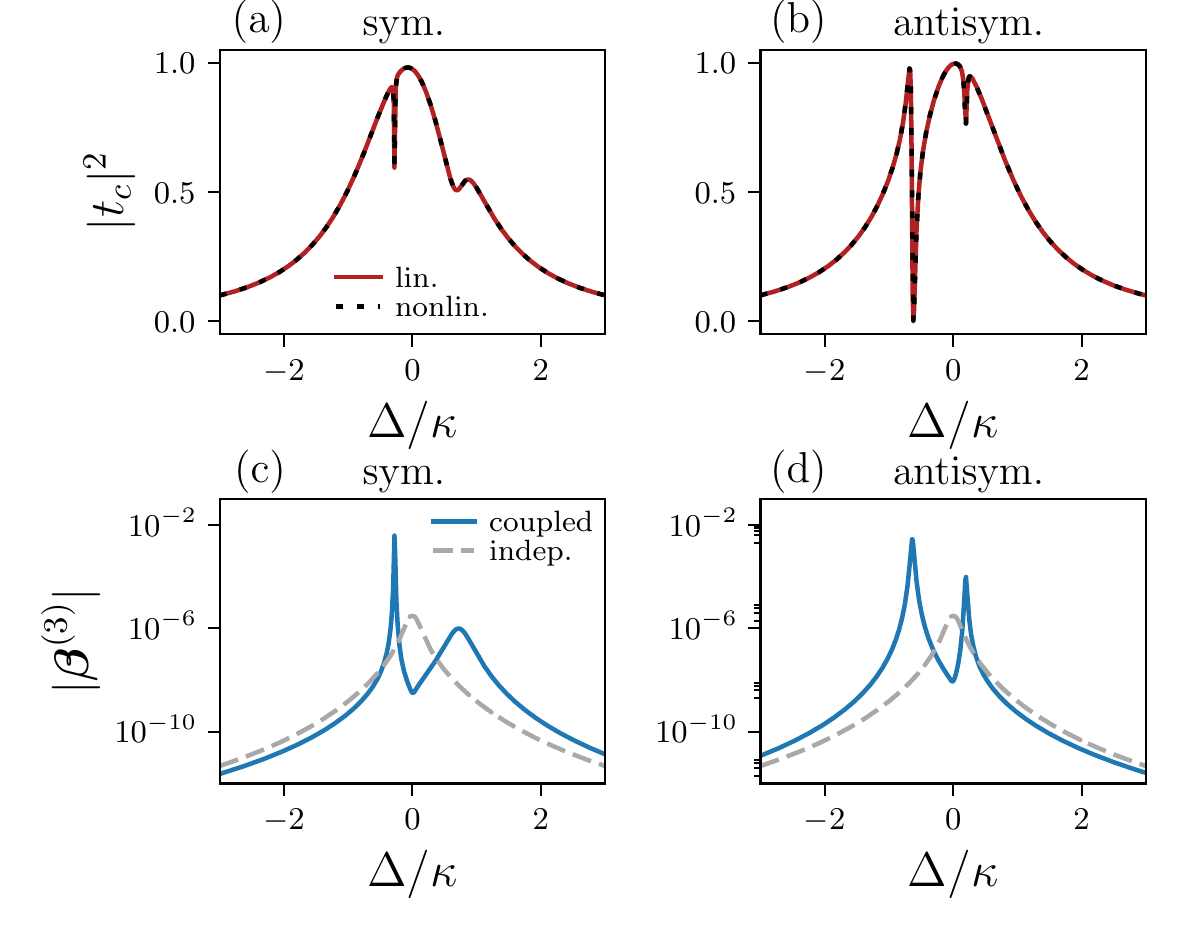}
\caption{\emph{Kerr nonlinearity for coupled quantum emitters}. Plot of transmission functions in the linear and nonlinear approximations for (a) symmetric and (b) asymmetric excitation of four equally spaced emitters at distance $d=0.07\lambda\ts{e}$. The corresponding nonlinearity is plotted in (c) for symmetric and (d) for asymmetric addressing compared to the independent emitter case (dashed line obtained setting $d\gg\lambda\ts{e}$). The remaining parameters here are $g=\kappa/10$, $\gamma=\kappa/20$, and $\eta=\kappa/100$. Note that no frequency matching was assumed, such that $\omega\ts{e}=\omega\ts{c}$ and a scan of the laser frequency will consequently hit all the collective states ($\Delta\ts{c}=\Delta\ts{e}=\Delta$), producing a set of four antiresonances.}
\label{fig8}
\end{figure}

\subsection{Collective nonlinear effects}

The vector of individual classical dipoles can be expressed in steady state as
\begin{align}
\pmb{\beta}^{(1)} = -i\eta[(\kappa-i\Delta\ts{c})(i\boldsymbol{\Omega}+\boldsymbol{\Gamma}-i\Delta\ts{e}\mathbbm{1}) + \textbf{G}\textbf{G}^\top]^{-1} \textbf{G}.
\end{align}
We then find the next order correction, similarly to the single-emitter case, by writing $\pmb{\beta}\approx \pmb{\beta}^{(1)}+ \pmb{\beta}^{(3)}$ (see Appendix~\ref{sec:App.Nonlinear-Coll} for details) and obtain a compact expression
\begin{align}
\label{beta-3}
\pmb{\beta}^{(3)} = & \frac{2i\eta \sum_{j=1}^{N}\textbf{P}_{j}\pmb{\beta}^{(1)}{\pmb{\beta}^{(1)}}^{\dagger}\textbf{P}_{j}}{(\kappa-i\Delta\ts{c})(i\boldsymbol{\Omega}+\boldsymbol{\Gamma}-i\Delta\ts{e}\mathbbm{1}) + \textbf{G}\textbf{G}^\top}\\
& \times \left(\textbf{G} - \frac{i}{\eta}\left[(\kappa - i\Delta_\text{c})(i\boldsymbol{\Omega}+\tilde{\boldsymbol{\Gamma}})+\textbf{G}\textbf{G}^{\top}\right]\boldsymbol{\beta}^{(1)}\right),
\notag
\end{align}
where $\tilde{\boldsymbol{\Gamma}} = \boldsymbol{\Gamma}-\gamma\mathbbm{1}$ and $\textbf{P}_j$ is the projector on the $j$th unit vector. This describes the collective cooperative Kerr effect where the induced nonlinear polarization of each emitter in the ensemble depends on the collective response of all the other emitters. As a basis for comparison, we estimate that for independent emitters, illuminated symmetrically, the maximum linear response per emitter (they all respond equally to the excitation) is $\beta^{(1)}_j(\Delta=0)=iC\ts{eff}/(1+C\ts{eff}) \eta/(N g)$. In particular, for $N=1$, one would have $\beta^{(1)}(\Delta=0)=iC/(1+C) \eta/g$. Notice that an increase in $N$ leads to an increase in the cooperativity such that the factor $C\ts{eff}/(1+C\ts{eff})$ increases but eventually saturates at unity. The other factor decreases with $N$ such that, in the many-emitter limit, the per-emitter nonlinearity decreases. Emitter-emitter coupling, on the other hand, can strongly modify the width of the antiresonance and consequently strongly enhance the overall nonlinearity.

This is illustrated in~\fref{fig8}. We consider a system of four coupled emitters exhibiting four collective states with energies approximately given by Eq.~\eqref{m_energies}. Three of these states are subradiant while the superradiant state is energetically located at the frequency $\omega_e+2\Omega\ts{12}\cos(\pi/5)$ on the right of the cavity resonance. We drive the system either symmetrically with $\textbf{G}=(g,g,g,g)^{\top}$ or asymmetrically with $\textbf{G}=(g,-g,g,-g)^{\top}$. On the left side, in ~\fsref{fig8}(a) and \ref{fig8}(c), the symmetric addressing partially overlaps with one subradiant state and one superradiant state. The coupling to the other two states is negligible. At the point where the laser fits the displaced collective states, the collective nonlinearity (shown as the norm of the $\pmb{\beta}^{(3)}$ vector) exhibits a large enhancement. The superradiant antiresonance, on the other hand, shows a decrease from the independent emitter's maximum nonlinearity (dashed line evaluated at the origin). Notice that in the independent case the four collective states have the same degenerate energy equal to $\omega\ts{e}$. For asymmetric driving, the laser encounters two subradiant collective states and shows the corresponding enhancement in the nonlinearity.

\begin{figure}[t]
\includegraphics[width=\columnwidth]{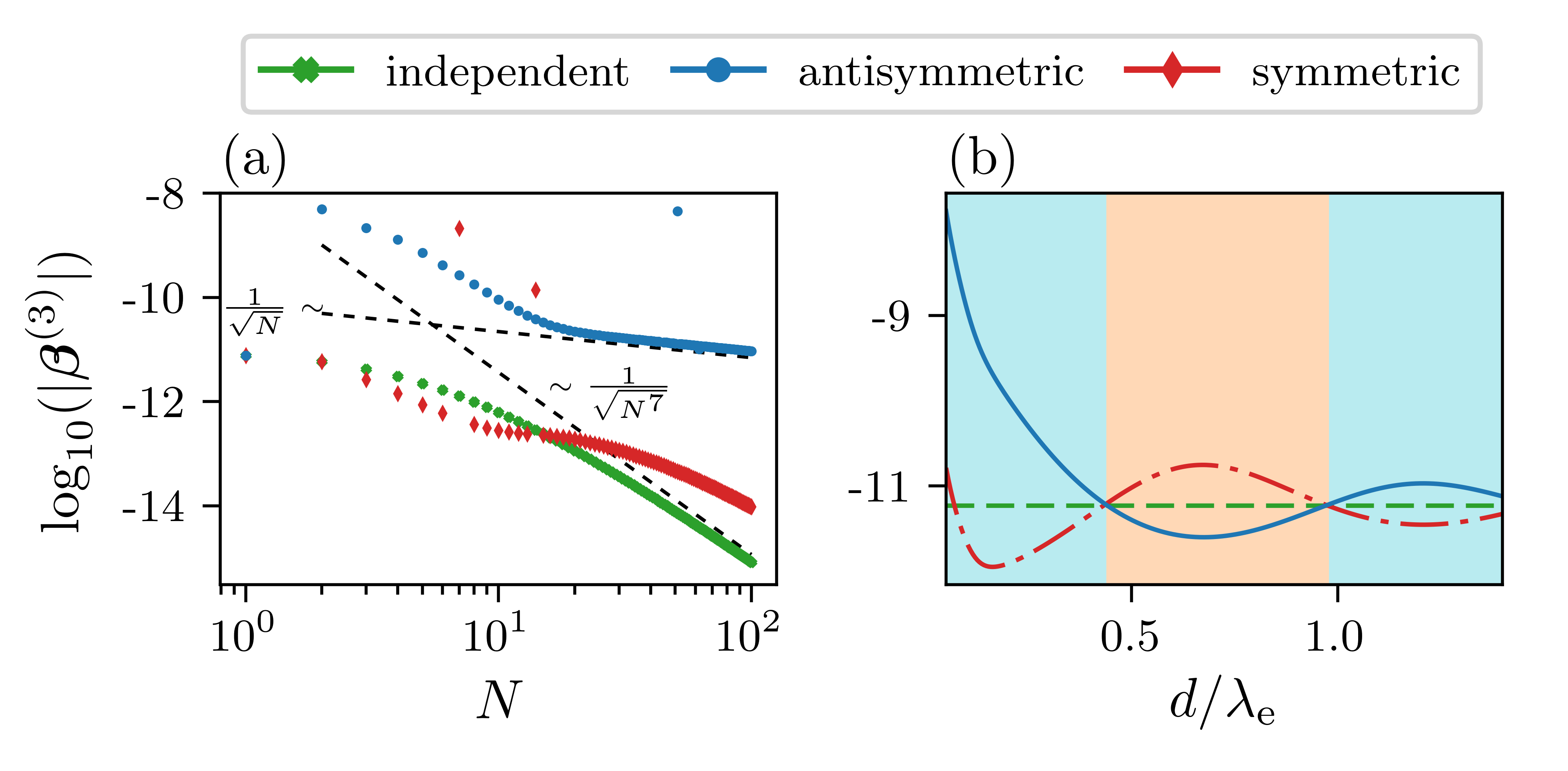}
\caption{\emph{Behavior of the Kerr nonlinearity}. (a) The nonlinearity as a function of $N$ for a chain with separation $d=0.07\lambda\ts{e}$. We observe different scaling laws in the limit where $N\gg 1$. While for independent emitters the nonlinearity attains the scaling with $1/\sqrt{N^7}$, in the case of asymmetric driving (subradiance) a much more robust scaling with $1/\sqrt{N}$ is found. Even for symmetrically coupled emitters (superradiance) the nonlinearity remains larger due to the presence of collective shifts. (b) The two-emitter nonlinearity as a function of the inter-particle distance $d$. The light blue area highlights where the nonlinearity for asymmetric addressing (blue, solid line) is larger than for symmetric addressing (red, dash-dotted line). The orange region highlights the opposite case. The regions coincide with $\gamma_{12}>0$ (blue) and $\gamma_{12}<0$ (orange, $0.5\lambda\ts{e}\leq d \leq \lambda\ts{e}$). Note that at extremely small distances the energetic shift $\Omega_{12}$ is so dominant that even under symmetric addressing the nonlinearity surpasses the independent one, despite the emitters being almost perfectly superradiant. Note, that we chose a small driving strength, $\eta=10^{-4}\kappa$, in order to ensure that the excited state population remains sufficiently small. This is why at a first glance the nonlinearity appears to be a lot smaller than the one shown in \fref{fig8}. The remaining parameters were $\gamma=\kappa/20$, $g=\kappa/10$.}
\label{fig9}
\end{figure}

In order to further investigate the physics here, we can look at two special cases. First, consider an ensemble of independent emitters all of which couple equally to the cavity mode. At resonance, we find that the magnitude of the Kerr nonlinearity is
\begin{align}
|\boldsymbol{\beta}^{(3)}(\Delta=0)| &= \frac{2\eta^3}{N}\sqrt{\frac{(NC)^3}{\gamma^3\left(1 + NC\right)^8\kappa^3}}.
\end{align}
In the limit of many emitters, $N\gg 1$, one can see that the nonlinearity at resonance scales as $1/\sqrt{N^7}$. This is expected as in this limit the ensemble of emitters more closely resembles a harmonic oscillator making the entire system linear. The situation is much less trivial for coupled emitters. The scaling of the collective Kerr nonlinearity with the number of emitters is shown in \fref{fig9}(a). There, it can be seen that depending on the symmetry of the coupling to the cavity mode, the scaling down with $N$ is drastically different. Namely, we find that even under symmetric addressing which leads to superradiance, the nonlinearity is larger than for uncoupled emitters. As shown below, this is due to the presence of collective shifts. Eventually, though, the nonlinearity attains a scaling close to the independent case in the limit of many emitters. Asymmetric addressing, on the other hand, leads to a completely different behavior. Since not only collective shifts are present, but also the linewidth is reduced due to subradiance, the resulting Kerr nonlinearity is much more robust; i.e., it scales down with $1/\sqrt{N}$.

The second special case we want to look at is the smallest collective system, which consists of two emitters only. If we consider symmetric (+) and asymmetric (-) addressing, i.e. $\textbf{G}=(g,\pm g)^T$, the collective shifts amount to $\pm \Omega_{12}$. Matching the cavity frequency to this collective resonance, we find the Kerr nonlinearity at resonance ($\Delta=\Delta\ts{c}=\Delta\ts{e} \mp\Omega_{12}$)
\begin{align}\label{eq:beta-3-coll}
|\boldsymbol{\beta}^{(3)}(\Delta=0)| &= \sqrt{\frac{C\ts{eff}^3}{(1+C\ts{eff})^3}}\frac{\eta^3\sqrt{\left(\gamma^2 + \Omega_{12}^2\right)}}{\sqrt{\left[\gamma\ts{eff}\left(1 + C\ts{eff}\right)\right]^5\kappa^3}}.
\end{align}
The first term above is the same as in $\beta^{(1)}$. The second term, on the other hand, exhibits some peculiarities. Specifically, while it is inversely proportional to the width of the antiresonance $\gamma\ts{eff}(1+C\ts{eff})$, it also depends on the collective shifts $\Omega_{12}$. Making use of subradiance can significantly decrease the antiresonance width. However, it is eventually limited by the decay channel constituted by the cavity with the rate $Ng^2/\kappa$. The shifts, though, can still compensate the broadening by the cavity and increase the nonlinearity above the one exhibited by decoupled emitters. At extremely small distances, where the shifts start to diverge, this effect is so predominant that it can even compensate for an additional broadening of the antiresonance due to superradiance. This can be seen in \fref{fig9}(b), where the behavior of the two-emitter nonlinearity as a function of the inter-particle distance is depicted. Since the shifts also increase with the number of emitters, the same argument applies to the fact that the symmetric nonlinearity surpasses the independent one in \fref{fig9}(a). At moderate particle-particle separation, the shifts are not too large, such that the nonlinearity is essentially governed by the collective decay [see \fref{fig9}(b)]. As the mutual decay rate $\gamma_{12}$ changes sign, symmetric addressing and asymmetric addressing switch roles, such that the system becomes subradiant under symmetric addressing and subsequently is more nonlinear.

\section{Conclusions}
We have followed a quantum Langevin equations approach to the input-output problem of an optical cavity containing an ensemble of $N$ coupled quantum emitters. Linearization of operators around steady-state values in the weak excitation regime allows one to compactly write the evolution of quantum fluctuations and derive expectation values not only for two-operator products but products of any number of operators. In particular, we focused on describing the properties of the reflected and transmitted output field as well as of the detected fields (assuming a flat-window time integration). We have developed the formalism first for the case of a single quantum emitter coupled to an optical cavity applicable both in the Purcell and strong coupling regimes. We have then extended this formalism to the case of many emitters where numerical simulations become difficult and found the signature of collective cooperative behavior: $N$ emitters do not only imprint an $N$-times larger effect on the cavity field but far beyond this linear scaling. The results originally shown in Ref.~\cite{plankensteiner2017cavity} were extended. There, the cooperativity was shown to increase drastically with $N$ when asymmetric, energetically matched excitation schemes were employed to prepare collective subradiant states. The formalism developed allows one to go beyond the classical problem and describe quantum properties of the output field. Moreover, the same \textit{cooperative collective} effects, stemming from the coupling among emitters, lead to a strong enhancement of the non-linear response of the cavity-embedded ensemble around specific antiresonances. A more detailed study of such effects (which are promising for nonlinear quantum optics applications) will be tackled in a future publication aiming at deriving precise conditions for the antiresonance points where the effect could be optimized even for distances achievable by optical lattice trapping techniques.

\begin{acknowledgments}
We acknowledge financial support from the Austrian Science Fund (FWF) within the Innsbruck Doctoral School DK-ALM: W1259-N27 (D.~P.) and Project No. P29318-N27 (H.~R.), from the Max Planck Society (M.~R., C.~S., and C.~G.) and from the German Federal Ministry of Education and Research, co-funded by the European Commission (Project RouTe), Project No. 13N14839 within the research program "Photonik Forschung Deutschland" (C.~S. and C.~G.).

The graphs were produced with the open-source plotting library Matplotlib \cite{hunter2007matplotlib} and some numerical calculations were carried out with the QuantumOptics.jl toolbox \cite{kramer2018quantumoptics}.
\end{acknowledgments}

\renewcommand{\theequation}{\Alph{section}.\arabic{equation}}

\section{Collective noise in quantum Langevin equations}
\label{sec:App.QLEs}
For a system operator $\hat{O}$, for each individual Lindblad collapse operator $\hat{c}$ acting at rate $\gamma\ts{c}$ and with associated input noise $\hat{c}\ts{in}$ one can derive the QLE including the noise terms as~\cite{gardiner2004quantum}
\begin{align}
\label{diagonal_HLE}
\dot{\hat{O}}=\frac{i}{\hbar}[H,\hat{O}]-[\hat{O},\hat{c}^\dagger]\left\{\gamma\ts{c} \hat{c}+\sqrt{2\gamma\ts{c}}\hat{c}\ts{in}\right\}+
\\
+\left\{\gamma\ts{c} \hat{c}^\dagger +\sqrt{2\gamma\ts{c}}\hat{c}\ts{in}^\dagger\right\}[\hat{O},\hat{c}]
\notag
\end{align}
However, collective incoherent dynamics represented by nondiagonal Liouvillian terms cannot be directly cast into Langevin equations. One instead has to first write the total decay term as a sum of independent decay channels. This is achieved by a basis transformation with the matrix $\textbf{T}$ (such that $\textbf{T}^{-1}=\textbf{T}^\top$) that diagonalizes the decay rate matrix $\boldsymbol{\Gamma}$,
\begin{align}
\text{diag}\left(\lambda_1,...,\lambda_N\right) &= \textbf{T}^\top\boldsymbol{\Gamma}\textbf{T},
\end{align}
where $\lambda_j$ is the $j$th eigenvalue of the decay matrix. Defining a set of damping operators
\begin{align}\label{Pi_op}
\hat \Pi_j &:= \sum_k \left(\textbf{T}^\top\right)_{jk}\hat S_k,
\end{align}
we may write \cite{ostermann2014protected}
\begin{align}
\mathcal{L}_\text{e}[\rho] &= \sum_i\lambda_i\left(2\hat \Pi_i\rho\hat \Pi_i^\dagger - \hat \Pi_i^\dagger \hat \Pi_i \rho - \rho\hat \Pi_i^\dagger \hat \Pi_i\right).
\end{align}
Obviously, this Lindblad term is diagonal and hence the QLE may be cast into the form given by Eq.~\eqref{diagonal_HLE}. The input noise terms of the emitter operators $\hat \sigma_{i,\text{in}}$ follow the transformation rules given by Eq.~\eqref{Pi_op}. Transforming the QLE for any emitter operator $\hat{O}$ back into the nondiagonal form gives the usual terms for the deterministic parts. For the noise terms, however, we have
\begin{align}
\sum_j[\hat{O},\hat \Pi_j^\dagger]\sqrt{\lambda_j}\hat \Pi_{j,\text{in}} &= \sqrt{\gamma}\sum_j[\hat{O},\hat S_j^\dagger]\xi_j(t),
\\
\sum_j[\hat{O},\hat \Pi_j]\sqrt{\lambda_j}\Pi_{j,\text{in}}^\dagger &= \sqrt{\gamma}\sum_j[\hat{O},\hat S_j]\xi_j^\dag(t),
\end{align}
where we have implicitly defined our correlated emitter noise terms $\xi_j$ as
\begin{align}
\xi_j(t) &:= \sum_{k,l} \sqrt{\frac{\lambda_k}{\gamma}}\textbf{T}_{jk}(\textbf{T}^{-1})_{kl}\hat \sigma_{l,\text{in}}^-.
\end{align}
Hence, the QLE for any emitter operator $\hat{O}$ is
\begin{align}
\dot{\hat{O}} &= i[H,\hat{O}] - \sum_{ij}[\hat{O},\hat S_i^\dagger]\left(\gamma_{ij}\hat S_j + \delta_{ij}\sqrt{2\gamma}\xi_i(t)\right) +
\notag \\
&+ \sum_{ij} \left(\gamma_{ij}\hat S_j^\dagger +\delta_{ij} \sqrt{2\gamma}\xi_i^\dag(t)\right)[\hat{O},\hat S_i],
\end{align}
with the spatially correlated white noise $\xi_i$. From the definition of the noise it is straightforward to show that
\begin{align}\label{app:noise-comm}
[\xi_i(t),\xi_j^\dag(t')] &= h_{ij}\delta(t-t').
\end{align}

In order to evaluate the correlation functions of the modified collective input noise terms,
\begin{align}
\bar{\xi}_j &= \hat S_j^z(t)\xi_j(t),
\\
\bar{\xi}_j^z(t) &= 2\left(\hat S_j^\dag(t)\xi_j(t) + \xi_j^\dag(t)\hat S_j(t)\right),
\end{align}
we need to consider the commutation relations of the system operators with the input noise terms $\xi_j(t)$. To this end, we write the collective input-output relation~\cite{gardiner2004quantum} which is straightforward from the diagonal form of the QLEs,
\begin{align} \label{app:coll-in-out}
\hat \Pi_{j,\text{in}}(t) + \hat \Pi_{j,\text{out}}(t) &= \sqrt{2\lambda_j}\hat \Pi_j(t).
\end{align}
Because of causality, it is clear that for any system operator $\hat{O}$,
\begin{align}
[\hat{O}(t), \hat \Pi_{j,\text{in}}(t')] &= 0, \quad t' > t;
\end{align}
i.e., the system does not depend on future input noise. For the output, we can invert this reasoning such that $\hat \Pi_{j,\text{out}}(t')$ commutes with $\hat{O}(t)$ if $t'<t$. Using these findings in combination with the input-output relation from Eq.~\eqref{app:coll-in-out}, we obtain~\cite{gardiner2004quantum}
\begin{align}
[\hat{O}(t), \hat \Pi_{j,\text{in}}(t')] &= \Theta(t-t')\sqrt{2\lambda_j}[\hat{O}(t),\hat \Pi_j(t')],
\end{align}
where we defined $\Theta$ as the step function with the half-maximum convention, i.e., $\Theta(0)=1/2$. Using the transformation between the diagonal operators and the correlated noise operators, $\xi_k(t) = \sum_j \sqrt{\lambda_j/\gamma}\textbf{T}_{kj}\hat \Pi_{j,\text{in}}(t)$, we find the commutation relations of a system operator with the correlated input noise,
\begin{align}\label{app:sys-noise-comm}
[\hat{O}(t),\xi_k(t')] &= \Theta(t-t')\sqrt{2\gamma}\sum_lh_{kl}[\hat{O}(t),\hat S_l(t')].
\end{align}
Using this, we can compute the correlation functions of the modified input noise operators,
\begin{align}
\braket{\bar{\xi}_j(t)\bar{\xi}^\dag_k(t')} &= h_{jk}\delta(t-t')\braket{\hat S_j^z(t)\hat S_k^z(t')} +
\notag \\
&+ \braket{\hat S_j^z(t)\xi_k^\dag(t')\xi_j(t)\hat S_k^z(t')},
\end{align}
where we used the commutation relation from Eq.~\eqref{app:noise-comm}. Using the commutation rules from Eq.~\eqref{app:sys-noise-comm}, one can show that the second term is proportional to $\Theta(t-t')\Theta(t'-t)$. Thus, it is only finite if $t=t'$ and does not contribute as a distribution. In other words, an integral over any time interval (such as our detection window) of this term vanishes. We can therefore neglect this term and arrive at
\begin{align}
\braket{\bar{\xi}_j(t)\bar{\xi}^\dag_k(t')} &= h_{jk}\delta(t-t')\braket{\hat S_j^z(t)\hat S_k^z(t')},
\end{align}
which in the linearized regime yields Eq.~\eqref{noise-corr-1}.

Proceeding, we have
\begin{align}
\braket{\bar{\xi}_j^z(t)\bar{\xi}_k^z(t')} &= 4\Big[h_{jk}\delta(t-t')\braket{\hat S_j^\dag(t)\hat S_k(t')} +
\notag \\
&+ \braket{\hat S_j^\dag(t)\xi_k^\dag(t')\xi_j(t)\hat S_k(t')}\Big].
\end{align}
Since $\xi_j(t)$ and $\hat S_k(t')$ commute if $t \geq t'$ and $\xi_j(t)$ applied to the right vanishes, the second term in the above expression is zero for $t\geq t'$. The same reasoning applies to $\xi_k^\dag(t')$ and $\hat S_j^\dag(t)$ for $t \leq t'$, such that we have
\begin{align}
\braket{\bar{\xi}_j^z(t)\bar{\xi}_k^z(t')} &= 4h_{jk}\delta(t-t')\braket{\hat S_j^\dag(t)\hat S_k(t')},
\end{align}
which after linearizing gives the expression in Eq.~\eqref{noise-corr-2}. Finally, we can use the same line of argument to derive the correlation function $\braket{\bar{\xi}_j(t)\bar{\xi}_k^z(t')}$ in Eq.~\eqref{noise-corr-end}.

\section{Steady-state Lyapunov equation}
\label{sec:App.Lyapunov}
The general solution for a system of $N$ linearly coupled QLEs with constant coefficients reads
\begin{align}
\textbf{v}(t) &= e^{\textbf{M}t} \textbf{v}(0) +\int_0^t dt' e^{\textbf{M}(t-t')}\textbf{N}\textbf{v}\ts{in}(t').
\end{align}
When the real part of all eigenvalues of the drift matrix \textbf{M} are negative, the system is stable and goes towards a steady state where $e^{\textbf{M}t}$ vanishes and the transient solution (containing information about the initial state) plays no role. In such a case, for times $t$ large enough such that steady state is already reached, one can define (a time-independent) correlation matrix $\textbf{V} = \braket{\textbf{v}(t)\textbf{v}^\top(t)}$ that is easily constructed with the steady-state solution only,

\begin{align}\label{covariance_1}
&\textbf{V} = \int_0^t dt' e^{\textbf{M}(t-t')} \textbf{D} e^{\textbf{M}^T(t-t')},
\notag
\end{align}
where we have use $\braket{\textbf{v}\ts{in}(t')\textbf{v}\ts{in}(t'')^\top}=\textbf{C}\delta(t'-t'')$ and defined the diffusion matrix as $\textbf{D}=\textbf{N}\textbf{C}\textbf{N}^T$. The expression for \textbf{C} can be computed from the input correlations and the results is listed in the main text as Eq.~\eqref{defC}. We can then derive a Lyuapunov equation for the covariance matrix using integration by parts
\begin{align}
\textbf{M}\textbf{V} &+ \textbf{V}\textbf{M}^\top = \int_0^t dt'\textbf{M}e^{\textbf{M}(t-t')} \textbf{D} e^{\textbf{M}^T(t-t')} + V\textbf{M}^\top =
\notag \\
&= e^{\textbf{M}(t-t')}\textbf{D}e^{\textbf{M}^\top (t-t')}\Big|_{t'=0}^t - \textbf{V}\textbf{M}^\top + \textbf{V}\textbf{M}^\top =
\notag \\
&= -\textbf{D}
\end{align}
Notice that one can write a similar equation for the symmetrized covariance matrix defined as $V_{ij} = \left(\braket{v_iv_j} + \braket{v_jv_i}\right)/2$, by a simple replacement of the diffusion matrix with the symmetrized one $\textbf{D} = \textbf{N}\left(\textbf{C} + \textbf{C}^\top\right)\textbf{N}^\top/2$.

\section{Fourier analysis of the output and the detected signal}
\label{sec:App.Fourier}

\subsection{Output spectrum}
We define the Fourier transform for an arbitrary operator $\hat{O}(t)$ as
\begin{align}
\hat{O}(t) &= \frac{1}{\sqrt{2\pi}}\int_{-\infty}^{\infty} d\omega e^{i\omega t}\hat{O}(\omega),
\end{align} which we employ to transform the linear set of differential equations to a set of coupled equations,
\begin{align}
i\omega\textbf{v}(\omega) &= \textbf{M}\textbf{v}(\omega) + \textbf{N}\textbf{v}\ts{in}(\omega).
\end{align}
This allows us to express the intracavity quantum fluctuations in terms of the input noise as
\begin{align}
\label{v_cav_fourier}
\textbf{v}(\omega) &= \left(i\omega \mathbbm{1} - \textbf{M}\right)^{-1}\textbf{N}\textbf{v}\ts{in}(\omega),
\end{align}
Furthermore, using input-output relations in the time domain
\begin{align}
\textbf{v}\ts{out}(t) &= \textbf{N}^\top\textbf{v}(t) - \textbf{v}\ts{in}(t),
\end{align}
allows us to connect the output to the input as a simple matrix multiplication,
\begin{align}
\textbf{v}\ts{out}(\omega) &= \textbf{F}(\omega)\textbf{v}\ts{in}(\omega),
\end{align}
where $\textbf{F}(\omega) = [\textbf{N}^{\top}\left(i\omega \mathbbm{1} - \textbf{M}\right)^{-1}\textbf{N} - \mathbbm{1}]$. In the frequency space, the response of the system ensures the preservation of $\delta$ correlations,
\begin{align}
\braket{\textbf{v}\ts{out}(\omega)\textbf{v}^\top \ts{out}(\omega')}=\pmb{\cal{S}}\ts{out}(\omega)\delta(\omega+\omega').
\end{align}
The system response for two-operator correlations is completely encoded in the spectrum matrix given by
\begin{align}
\pmb{\cal{S}}\ts{out}(\omega) =\textbf{F}(\omega)\textbf{C}\textbf{F}^{\top}(-\omega).
\end{align}

\subsection{Time-integrated correlations}
The output can be used directly to compute correlations for time-integrated operators at equal times,
\begin{align}
\textbf{V}\ts{det}(t) &= \frac{1}{2T}\int_{t - T}^{t + T}dt'\int_{t - T}^{t + T}dt''\braket{\textbf{v}\ts{out}(t')\textbf{v}\ts{out}(t'')^\top}.
\end{align}
We then expand operators in terms of their Fourier components and use the $\delta$ correlations in the frequency space. This leads to the evaluation of the following integral:
\begin{align}
\int_{t-T}^{t+T}dt' \int_{t-T}^{t+T}dt''e^{i\omega t'}e^{-i\omega t''} =\left[ \frac{2\sin{\omega T}}{\omega}\right]^2.
\end{align}
For sufficiently long detection times (much longer than the characteristic bandwidth of the spectrum matrix), the sinc function picks only the zero-frequency component (since we are in a rotating frame, this corresponds to the laser frequency),
\begin{align}
\textbf{V}\ts{det} &= \frac{1}{\pi T}\int_{-\infty}^{\infty}d\omega \left[\frac {\sin^2{\omega T}}{\omega^2}\right]\pmb{\cal{S}}\ts{out}(\omega) \approx \pmb{\cal{S}}\ts{out}(0).
\end{align}

\subsection{Resolving four-point correlations}
Let us generally write expectation values for any combinations of detected operators as ${\cal{R}}_{i_1 i_2 i_3 i_4}=\langle v^{i_1}\ts{det} v^{i_2}\ts{det} v^{i_3}\ts{det} v^{i_4}\ts{det} \rangle$.
We then connect these expectation values to the output operator combinations for which Isserlis' theorem applies, allowing one to express any product of operators in sums of all different products of two-point correlations as
\begin{widetext}
\begin{align}
\langle v^{i_1}\ts{out}(\omega_1)v^{i_2}\ts{out}(\omega_2) v^{i_3}\ts{out}(\omega_3) v^{i_4}\ts{out}(\omega_4) \rangle =& \quad {\cal{S}}^{_{i_1 i_2}}\ts{out}(\omega_1){\cal{S}}^{_{i_3 i_4}}\ts{out}(\omega_3)\delta(\omega_1+\omega_2)\delta(\omega_3+\omega_4)
\\& +{\cal{S}}^{_{i_1 i_3}}\ts{out}(\omega_1){\cal{S}}^{_{i_2 i_4}}\ts{out}(\omega_2)\delta(\omega_1+\omega_3)\delta(\omega_2+\omega_4)\nonumber
\\& +{\cal{S}}^{_{i_1 i_4}}\ts{out}(\omega_1){\cal{S}}^{_{i_2 i_3}}\ts{out}(\omega_2)\delta(\omega_1+\omega_4)\delta(\omega_2+\omega_3).\nonumber
\end{align}
\end{widetext}
The integration over the detection time window will, of course, again give rise to the sinc functions in the integrand; in the long detection time limit we then write the general expression
\begin{align}
{\cal{R}}_{i_1 i_2 i_3 i_4} ={\cal{S}}^{_{i_1 i_2}}\ts{out}(0){\cal{S}}^{_{i_3 i_4}}\ts{out}(0)&+{\cal{S}}^{_{i_1 i_3}}\ts{out}(0){\cal{S}}^{_{i_2 i_4}}\ts{out}(0)
\\ &+{\cal{S}}^{_{i_1 i_4}}\ts{out}(0){\cal{S}}^{_{i_2 i_3}}\ts{out}(0),\nonumber
\end{align}
which we can make use of to evaluate any four-point operator correlations. For example, we can evaluate the first term in the left-hand-side of Eq.~\eqref{variance} as
\begin{align}
\langle b^\dagger\ts{det}b\ts{det}b^\dagger\ts{det}b\ts{det} \rangle &= {\cal{S}}^{43}\ts{out}(0){\cal{S}}^{43}\ts{out}(0)+{\cal{S}}^{44}\ts{out}(0){\cal{S}}^{33}\ts{out}(0)+
\notag \\
&+{\cal{S}}^{43}\ts{out}(0){\cal{S}}^{34}\ts{out}(0).
\end{align}

\section{Free space spatial field distribution}
\label{E-field-derivation}
In order to express the electric field amplitude as function of the emitter operators $\hat S_i$, we follow an approach along the lines of Ref.~\cite{lehmberg1970radiation}. Recall that the Heisenberg equation for the photon annihilation operator of a mode with wave vector $\textbf{k}$ and polarization $\lambda$ coupled to $N$ identical quantum emitters is
\begin{align}
\dot{\hat{A}}_{\textbf{k},\lambda}(t) &= -ig_{\textbf{k},\lambda}\sum_j\hat{S}_j(t)e^{-i\textbf{k}\cdot \textbf{r}_j} e^{i(\omega_k - \omega\ts{e})t}.
\end{align}
Here, $g_{\textbf{k},\lambda} = \sqrt{\omega_k/(2\hbar\epsilon_0 V}\textbf{e}_{\textbf{k},\lambda}\cdot \boldsymbol{\mu}$ is the dipole interaction between the mode and the emitter with dipole moment $\boldsymbol{\mu}$. Note, that the equation above is already written in a frame rotating at $\omega_k - \omega\ts{e}$. The formal solution of the above equation is
\begin{align}
\hat{A}_{\textbf{k},\lambda}(t) &= \hat{A}_{\textbf{k},\lambda}(0)  +
\\
&- ig_{\textbf{k},\lambda}\sum_j\int_0^t dt'\hat{S}_j(t')e^{-i\textbf{k}\cdot \textbf{r}_j}e^{i\left(\omega_k - \omega\ts{e}\right)t'}
\notag
\end{align}
The initial value above corresponds to the input of the mode. Assuming that the modes surrounding the emitters are in vacuum, this term does not contribute to any averages. We therefore drop it in the following calculation.

Performing the Markov approximation, we can already see that it is possible to express the field annihilation operators at any time directly by the emitter operator at equal time.

Inserting the expression obtained for $\hat{A}_{\textbf{k},\lambda}$ results in the electric field amplitude
\begin{align}
\hat{\textbf{E}}^+(\textbf{r},t) &= -i\sum_j \hat{S}_j(t)\sum_{\textbf{k}}\textbf{f}(\textbf{k})\int_0^t dt' e^{-i\left(\omega_k - \omega\ts{e}\right)\left(t-t'\right)},
\end{align}
where
\begin{align}
\textbf{f}(\textbf{k}) &= \sum_\lambda \sqrt{\frac{\hbar\omega_k}{2\epsilon_0 V}}g_{\textbf{k},\lambda}\textbf{e}_{\textbf{k},\lambda}e^{i\textbf{k}\cdot(\textbf{r} - \textbf{r}_j)} =
\notag \\
&= \frac{\omega_k}{2\epsilon_0 V}\left(\boldsymbol{\mu} - \frac{\boldsymbol{\mu}\cdot\textbf{k}}{k^2}\textbf{k}\right)e^{i\textbf{k}\cdot(\textbf{r} - \textbf{r}_j)}.
\end{align}
In the last step, we exploited the liberty of choosing $\boldsymbol{\mu}$ to lie in the plane spanned by $\textbf{k}$ and $\textbf{e}_{\textbf{k}1}$ to resolve the sum over the polarization $\lambda=1,2$.

Since the set of free space modes is continuous, we can replace the sum over wave vectors by an integral
\begin{align}
\sum_\textbf{k} \to \frac{V}{(2\pi c)^3}\int d\omega_k \omega_k^2\int d\Omega_k,
\end{align}
where we have already written the integral in spherical coordinates in $\textbf{k}$ space. The part of the expression that has an angular dependence can be separated and the solid angle integral can be solved (for arbitrary $\textbf{r}$ and $\textbf{k}$),
\begin{align}
&\int d\Omega_k \left(\boldsymbol{\mu} - \frac{\boldsymbol{\mu}\cdot\textbf{k}}{k^2}\textbf{k}\right) e^{i\textbf{k}\cdot\textbf{r}} =
\notag \\
&= 2\pi\left(\boldsymbol{\mu} + \frac{(\boldsymbol{\mu}\cdot\nabla_\textbf{r})}{k^2}\nabla_\textbf{r}\right)\int_0^\pi d\theta \sin\theta e^{ikr\cos\theta}.
\end{align}
Here, we already solved the polar angle integral obtaining $2\pi$ and written the products with $\textbf{k}$ as derivatives of the exponential function. The remaining integral is straightforward to solve and the problem of solving the solid angle integral surmounts to applying the Nabla operator.

Inserting the result back into the electric field, we obtain
\begin{align}
\hat{\textbf{E}}^+(\textbf{r},t) &= \frac{-i\mu}{\epsilon_0(2\pi)^2c^3}\sum_j\hat{S}_j(t)\int d\omega_k\omega_k^3
\notag \\
&\int_0^t dt' e^{-i\left(\omega_k - \omega\ts{e}\right)\left(t-t'\right)}\times\sum_{i\in\{x,y,z\}} F_i(k|\textbf{r}-\textbf{r}_j|)\textbf{e}_i,
\end{align}
where $\textbf{e}_{x,y,z}$ is the respective unit vector in Cartesian coordinates and
\begin{widetext}
\begin{subequations}
\begin{align}
F_x(kr) &= -\cos\theta\sin\theta\cos\phi\left(\frac{\sin(kr)}{kr} + 3\frac{\cos(kr)}{(kr)^2} - 3\frac{\sin(kr)}{(kr)^3}\right),
\\
F_y(kr) &= -\cos\theta\sin\theta\sin\phi\left(\frac{\sin(kr)}{kr} + 3\frac{\cos(kr)}{(kr)^2} - 3\frac{\sin(kr)}{(kr)^3}\right),
\\
F_z(kr) &= \sin^2\theta\frac{\sin(kr)}{kr} + \left(1-3\cos^2\theta\right)\left(\frac{\cos(kr)}{(kr)^2} - \frac{\sin(kr)}{(kr)^3}\right).
\end{align}
\end{subequations}
\end{widetext}
In order to solve the time integral we make use of the Sokhotski formula,
\begin{align}
&\int d\omega_k\omega_k^3\int_0^t dt' e^{-i\left(\omega_k - \omega\ts{e}\right)\left(t-t'\right)} F_i(kr) =
\\
&= \int d\omega_k \omega_k^3\left(-i\mathcal{P}\frac{1}{\omega_k-\omega\ts{e}} + \pi\delta(\omega_k - \omega\ts{e})\right)F_i(kr),
\notag
\end{align}
where $\mathcal{P}$ denotes the principal value. The integral proportional to the $\delta$ distribution is straightforward to solve, while the principal value integrals require some more elaborate (yet standard) methods of complex contour integration.

Finally, the resulting electric field is given by
\begin{widetext}
\begin{align}
\label{E-field-result}
&\hat{\textbf{E}}^+(\textbf{r},t) = -i \frac{3\gamma}{4\mu}\sum_j S_j(t)\times\sum_{m\in\{x,y,z\}}\left( F_m(k\ts{e}|\textbf{r} - \textbf{r}_j|) - iG_m(k\ts{e}|\textbf{r}-\textbf{r}_j|)\right),
\end{align}
\end{widetext}
with
\begin{widetext}
\begin{subequations}
\begin{align}
G_x(kr) &= -\cos\theta\sin\theta\cos\phi\left(\frac{\cos(kr)}{kr} - 3\frac{\sin(kr)}{(kr)^2} - 3\frac{\cos(kr)}{(kr)^3}\right),
\\
G_y(kr) &= -\cos\theta\sin\theta\sin\phi\left(\frac{\cos(kr)}{kr} - 3\frac{\sin(kr)}{(kr)^2} - 3\frac{\cos(kr)}{(kr)^3}\right),
\\
G_z(kr) &= \sin^2\theta\frac{\cos(kr)}{kr} - \left(1-3\cos^2\theta\right)\left(\frac{\sin(kr)}{(kr)^2} + \frac{\cos(kr)}{(kr)^3}\right).
\end{align}
\end{subequations}
\end{widetext}
Note that one can easily calculate the dipole interaction with the field given by Eq.~\eqref{E-field-result}. This selects the component parallel to the dipole moment $\boldsymbol{\mu}$ (the $z$ component) and we obtain the effective emitter dipole-dipole interactions \cite{lehmberg1970radiation},
\begin{align}
\Omega_{ij} &= -\frac{3\gamma}{4}G_z\left(k\ts{e}|\textbf{r}_i - \textbf{r}_j|\right),
\\
\gamma_{ij} &= \gamma h_{ij} = \frac{3\gamma}{2}F_z\left(k\ts{e}|\textbf{r}_i - \textbf{r}_j|\right).
\end{align}

The intensity shown in \fref{fig4} is the average intensity at $t=0$ given by
\begin{align}
I(\textbf{r}) &= \braket{\hat{\textbf{E}}(\textbf{r})\cdot \hat{\textbf{E}}(\textbf{r})} = \braket{\hat{\textbf{E}}^+(\textbf{r})\cdot \hat{\textbf{E}}^-(\textbf{r})}.
\end{align}

\section{Correlation matrix for many emitters}
\label{N-emitters-matrices}
The definition of the many-emitter autocorrelation matrix as given in Eq.~\eqref{C_many-emitters} is rather condensed. Hence, here we specify once again what the matrix elements required to write this matrix down are. In essence, it boils down to the noise correlation functions given in Eqs. \eqref{noise-corr}. In particular, the $N\times N$ matrices used to define the overall autocorrelation matrix $\textbf{C}$ have the matrix elements
\begin{subequations}
\begin{align}
\braket{\bar{\xi}_j(t)\bar{\xi}_{k}^\dag(t')} &= \textbf{C}_{\beta\beta}^{jk}\delta(t-t'),
\\
\braket{\bar{\xi}_j^z(t)\bar{\xi}_{k}^z(t')} &= \textbf{C}_{zz}^{jk}\delta(t-t'),
\\
\braket{\bar{\xi}_j^z(t)\bar{\xi}_k^\dag(t')}  &= \textbf{C}_{z\beta}^{jk}\delta(t-t'),
\\
\braket{\bar{\xi}_j(t)\bar{\xi}_k^z(t')} &= \textbf{C}_{\beta z}^{jk}\delta(t-t').
\end{align}
\end{subequations}

\section{Nonlinear correction}
\label{sec:App.Nonlinear-Coll}
Starting from the QLEs for $N$ emitters which are given by
\begin{align}
\dot \alpha &= -\left(\kappa-i\Delta_\text{c}\right)\alpha + \eta - i\textbf{G}^{\top}\boldsymbol{\beta},
\\
\dot{\boldsymbol{\beta}} &= ((i\Delta\ts{e}-\gamma)\mathbbm{1} + \textbf{z}(i\boldsymbol{\Omega}+ \tilde{\boldsymbol{\Gamma}}))\boldsymbol{\beta} + i \textbf{z}\textbf{G}\alpha ,
\end{align}
where we define $\tilde{\boldsymbol{\Gamma}} = \boldsymbol{\Gamma} - \gamma\mathbbm{1}$, $\textbf{z} = \sum_{j=1}^{N} z_{j}\textbf{P}_{j}$, and $\textbf{P}_{j} = \textbf{e}_{j}\textbf{e}^{\top}_{j}$ ($\textbf{e}_{j}$ being a Cartesian basis vector). By employing the relation $z_{j} \approx 2|\beta_{j}|^2-1$, we obtain the equations
\begin{align}
\dot \alpha = & -\left(\kappa-i\Delta_\text{c}\right)\alpha + \eta - i\textbf{G}^{\top}\boldsymbol{\beta},
\\\nonumber
\label{App.Non.eq1}
\dot{\boldsymbol{\beta}} = & -(i\boldsymbol{\Omega}+ \boldsymbol{\Gamma}-i\Delta\ts{e})\boldsymbol{\beta} - i \textbf{G}\alpha +2|\boldsymbol{\beta}|^2(i\boldsymbol{\Omega}+ \tilde{\boldsymbol{\Gamma}})\boldsymbol{\beta}\\ &+ i2|\boldsymbol{\beta}|^2\textbf{G}\alpha,
\end{align}
with the matrix $|\boldsymbol{\beta}|^2 = \sum_{j=1}^{N}\textbf{P}_{j}\boldsymbol{\beta}\boldsymbol{\beta}^{\dagger}\textbf{P}_{j}$.
For the steady-state scenario with $\dot \alpha = 0$, we obtain $\alpha = (\eta-i\textbf{G}^{\top}\boldsymbol{\beta})/(\kappa-i\Delta_c)$. Substituting this into the steady-state equation ($\dot{\boldsymbol{\beta}} = 0$) of Eq.~\eqref{App.Non.eq1} results in
\begin{align}
\nonumber
0 = & -\left[(-i\Delta_\text{c}+\kappa)(-i \Delta\ts{e}\mathbbm{1} +i\boldsymbol{\Omega}+ \boldsymbol{\Gamma})+\textbf{G}\textbf{G}^{\top}\right]\boldsymbol{\beta} -i\textbf{G}\eta \\
\label{App.Non.eq2}
& + i2|\boldsymbol{\beta}|^2\textbf{G}\eta + 2|\boldsymbol{\beta}|^2\left[(i\Delta_\text{c}+\kappa)(i\boldsymbol{\Omega}+ \tilde{\boldsymbol{\Gamma}}) + \textbf{G}\textbf{G}^{\top}\right]\boldsymbol{\beta}.
\end{align}
With the linear solution being
\begin{align}
\label{App.Non.eq3}
\boldsymbol{\beta}^{(1)} = -i\eta\left[(\kappa-i\Delta_\text{c})(i\boldsymbol{\Omega}+ \boldsymbol{\Gamma}-i \Delta\ts{e}\mathbbm{1})+\textbf{G}\textbf{G}^{\top}\right]^{-1}\textbf{G},
\end{align}
the next order of correction can be found by introducing the perturbative ansatz $\boldsymbol{\beta} \approx \boldsymbol{\beta}^{(1)} + \boldsymbol{\beta}^{(3)}$ into Eq.~\eqref{App.Non.eq2} which leads to
\begin{widetext}
\begin{align}
\boldsymbol{\beta}^{(3)} = & 2\left[(\kappa-i\Delta_\text{c})(i\boldsymbol{\Omega}+ \boldsymbol{\Gamma}-i\Delta\ts{e})+\textbf{G}\textbf{G}^{\top}\right]^{-1} \times\left(\sum_{j=1}^{N}\textbf{P}_{j}\boldsymbol{\beta}^{(1)}\boldsymbol{\beta}^{(1)\dagger}\textbf{P}_{j}\right)\\
& \times \left(i\eta\textbf{G}+[(\kappa-i\Delta_\text{c})(i\boldsymbol{\Omega}+ \tilde{\boldsymbol{\Gamma}})+\textbf{G}\textbf{G}^{\top}]\boldsymbol{\beta}^{(1)}\right).
\nonumber
\end{align}
\end{widetext}
Here, we have ignored all terms with $O(\eta^4)$. The term $\boldsymbol{\beta}^{(3)}$ describes the collective Kerr nonlinearity of the $N$-emitter system. For $N=1$, this simplifies to $\beta^{(3)} = -2\beta^{(1)}|\beta^{(1)}|^{2}\left(1-i(g/\eta)\beta^{(1)}\right)$.

The modified transmission amplitude can be obtained from the relation $t  = (\kappa-i(\kappa/\eta)\textbf{G}^{\top}(\boldsymbol{\beta}^{(1)}+\boldsymbol{\beta}^{(3)}))/(\kappa-i\Delta_c)$. For a single emitter, we have
\begin{widetext}
\begin{align}
t\ts{c} = & t\ts{c}^{(1)}\left(1 + \frac{2g^4\eta^2}{\left((\gamma-i\Delta_{\mathrm{e}})(\kappa-i\Delta_{\mathrm{c}})+g^2\right)|(\gamma-i\Delta_{\mathrm{e}})(\kappa-i\Delta_{\mathrm{c}})+g^2|^2}\right) ,
\end{align}
\end{widetext}
where $t\ts{c}^{(1)} = \kappa/[(\kappa-i\Delta_{\text{c}})+g^2/(\gamma-i\Delta_{\text{e}})]$ is the result for the transmission in the linear case.

\end{document}